\documentclass{article}

\usepackage{arxiv}

\usepackage[utf8]{inputenc} % allow utf-8 input
\usepackage[T1]{fontenc}    % use 8-bit T1 fonts
\usepackage{hyperref}       % hyperlinks
\usepackage{url}            % simple URL typesetting
\usepackage{booktabs}       % professional-quality tables
\usepackage{array}
\usepackage{amsfonts}       % blackboard math symbols
\usepackage{nicefrac}       % compact symbols for 1/2, etc.
\usepackage{graphicx}
\usepackage{verbatim}
\usepackage{subcaption}% for subfigs
\usepackage{adjustbox}
\usepackage[numbers,sort&compress]{natbib}
\usepackage{doi}
\usepackage{amsmath}
\usepackage{amssymb}
\usepackage{algorithm}
\usepackage{algpseudocode}
\usepackage{xcolor}
\usepackage{mathtools}
\usepackage{fancyhdr}
\usepackage{amsthm}
\usepackage{cleveref}
\usepackage{bm}
\usepackage[mathscr]{euscript}
\usepackage{textcomp}

\DeclareMathAlphabet{\pazocal}{OMS}{zplm}{m}{n}

\usepackage{lscape}		
\usepackage{fancyvrb}

\newtheoremstyle{remarkstyle}  % Name
  {5pt}                        % Space above
  {5pt}                        % Space below
  {}                           % Body font (upright)
  {}                           % Indent amount
  {\bfseries}                  % Theorem head font (bold)
  {.}                          % Punctuation after theorem head
  { }                          % Space after theorem head
  {}                           % Theorem head spec (empty = `normal`)

\theoremstyle{remarkstyle}

\crefname{figure}{Fig.}{Figs.}
\Crefname{figure}{Figure}{Figures}
\crefname{algorithm}{Algorithm}{Algorithms}
\crefname{equation}{Eq.}{Eqs.}
\crefname{section}{Section}{Sections}

\title{
Neural Operator Representation of Granular Micromechanics-Based Failure Envelopes
}

\author{
Jinkyo Han$^{1}$, 
Payam Poorsolhjouy$^{3}$, 
Bahador Bahmani$^{1,2}$\thanks{Corresponding author: \texttt{bahador.bahmani@northwestern.edu}} \\
\\
$^{1}$Department of Mechanical Engineering, Northwestern University, Evanston, IL 60208, USA \\
$^{2}$Theoretical and Applied Mechanics, Northwestern University, Evanston, IL 60208, USA \\
$^{3}$Department of the Built Environment, Eindhoven University of Technology, 5612 AZ Eindhoven, Netherlands
}

\renewcommand{\shorttitle}{GMA-based Neural Operator}

\hypersetup{
    colorlinks=true,      % Enable colored links
    linkcolor=green,      % Color for internal links
    urlcolor=blue,        % Color for URLs
    citecolor=blue        % Color for citations
}

\date{}

\begin{document}
\maketitle

\vspace{-10pt}
\begin{abstract}
Micromechanics-based granular models are widely used to predict the failure behavior of porous and particulate materials, including concrete, soils, foams, and biological tissues. Although these models offer considerable flexibility through microstructural parametrization and statistical representation, their mapping to macroscopic responses, particularly failure envelopes, is implicit and requires costly nonlinear, non-smooth simulations, where each failure point is obtained by following a loading trajectory. This limitation is further amplified in inverse settings, where one seeks microstructure configurations that reproduce a target failure response. In this work, we propose a differentiable neural operator that learns the mapping from microstructure configurations to failure envelopes, enabling efficient forward prediction and inverse identification without repeated micromechanical simulations. To ensure mechanical admissibility, we incorporate a physics-informed training strategy that enforces convexity of the predicted envelopes, consistent with Drucker’s postulate, thereby eliminating potential non-physical artifacts. 
We also compare finite difference and automatic differentiation for evaluating the proposed regularization, and find that finite difference provides a favorable practical trade-off in the present DeepONet-based setting.
The operator is trained on failure envelopes represented as irregular point clouds, allowing learning from data sampled at heterogeneous resolutions. 
To further reduce computational cost, we introduce an active learning strategy that adaptively queries the micromechanical model in regions of high epistemic uncertainty. This leads to efficient exploration of the parameter space with fewer high-fidelity simulations. The versatility and performance of the method are demonstrated and benchmarked through several numerical examples.
\end{abstract}

% keywords can be removed

\section{Introduction}
The macroscopic failure of quasi-brittle materials, including rocks, concrete, and advanced ceramics, arises from the collective and competing microscale mechanisms of load-bearing, damage, and degradation \cite{haimson2010effect, Ma2016AGU, mourlas2023large}. To incorporate these effects into macroscopic descriptions, homogenization-based frameworks have been developed that represent the influence of the microstructure in an averaged sense.
Among these, orientation-based formulations such as the Microplane model \cite{bavzant1988microplane} and the Granular Micromechanics Approach (GMA) \cite{Cha1988} describe the macroscopic response through quantities defined over many orientations. While they differ in their physical interpretation and level of micromechanical detail, both these approaches avoid explicit realization of the full microstructure and instead obtain the macroscopic behavior through integration over orientations.
Variants of GMA have been successfully applied to a wide range of materials, including consolidated rocks, cemented sands, and cementitious composites \cite{Cha1988, chang1989, chang1990, liao2000, Suiker2001a,Suiker2001b,Suiker2004,misra2010,misra2014,misra2020granular,poorsolhjouy2017effect,bryant2022multiscale}.

Despite these strengths, GMA-based frameworks face two critical challenges. First, the calibration of microscopic constitutive relationships is non-trivial. Unlike explicit discrete simulations where physical laws like Hertzian contact may be directly applied \cite{qu2020hybrid}, the microscopic constitutive relationships in frameworks such as GMA and Microplane represent the response of microscopic constituents that are embedded within the material's microstructure. Consequently, these microscopic constitutive relationships often do not reflect the behavior of isolated features, and their design is largely based on intuitive understanding of the physical mechanisms involved \cite{bazant2000microplane}.
This becomes particularly evident when calibrating the microscopic constitutive relationships to achieve a target, perhaps experimentally derived, macroscopic failure envelope. 

\begin{figure}[htbp]
    \centering
    \begin{subfigure}{0.3\textwidth}
        \centering
        \includegraphics[width=\textwidth]{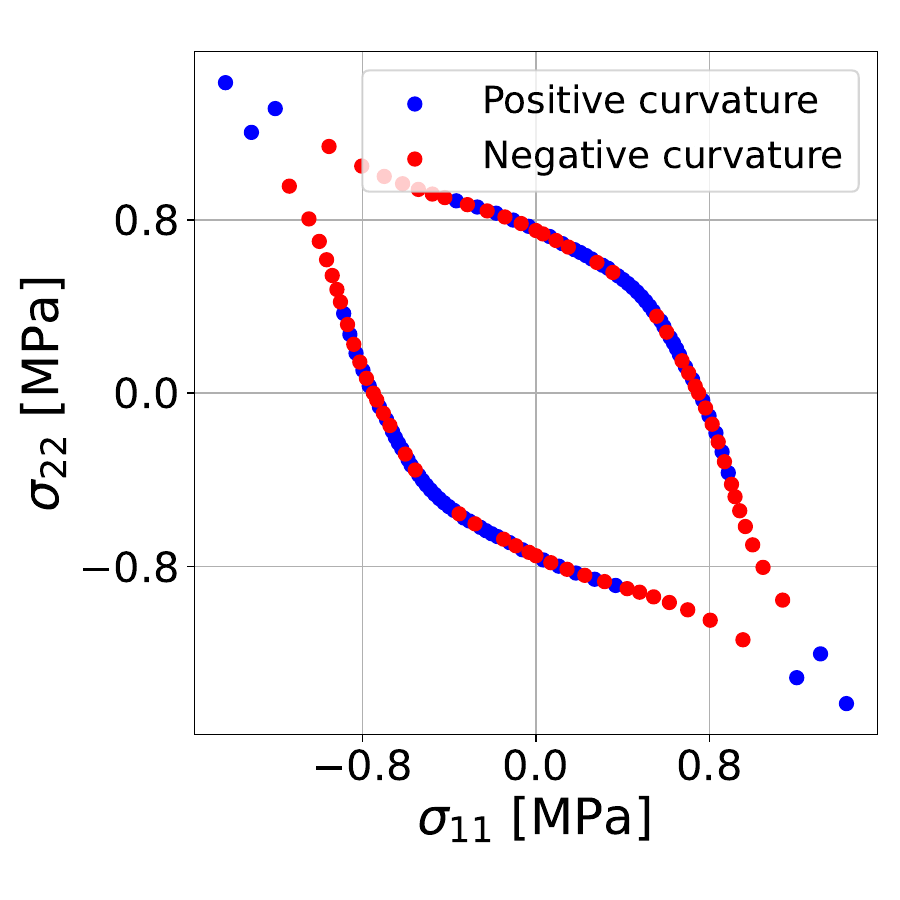}
        % \caption{}\label{fig:intro_nonconvex_a}
    \end{subfigure}
    % \begin{subfigure}{0.4\textwidth}
    %     \centering
    %     \includegraphics[width=\textwidth]{figures/intro_curvature_models_a.pdf}
    %     \caption{}\label{fig:intro_nonconvex_b}
    % \end{subfigure}
    \hfill
    \caption{
    An example in which the GMA model output exhibits local non-convex behavior. The pointwise curvature values are evaluated using a central difference scheme.
    }
    \label{fig:intro_nonconvex}
\end{figure}

Second, it has been observed that homogenization-based models derived from independent directional projections, such as GMA and the Microplane model, may produce failure envelopes that exhibit undesirable geometric features, including sharp vertices or spurious non-convexities \cite{caner2002vertex,misra2020granular}, as illustrated in 
% Fig.~\ref{fig:intro_nonconvex_a}.
\cref{fig:intro_nonconvex}.
While these features may emerge from the orientation-dependent softening of microscopic constituents, they may present challenges for macroscopic structural analysis. In particular, the presence of non-convex regions is inconsistent with Drucker’s postulate \cite{drucker1951fundamental,drucker1959definition}, which requires a convex failure surface to ensure stable material response.

While homogenization-based schemes are significantly more efficient than direct numerical simulations, they still require a considerable computational effort to construct the macroscopic failure envelope. In particular, the material point must be probed over a range of loading paths, and for each path, a nonlinear incremental analysis is performed to extract a single point on the failure surface \cite{poorsolhjouy2025rock}. This makes the generation of a full failure envelope computationally intensive. Moreover, the resulting mapping from micromechanical configuration to the failure envelope is neither explicit nor smooth, which poses challenges for gradient-based inverse identification.

To address these limitations, we propose a data-driven operator learning framework based on DeepOnet \cite{lu2021learning} that directly maps micromechanical configurations to the corresponding failure envelopes, bypassing the need for repeated directional loading simulations. By learning this mapping from data generated by the micromechanical model, the proposed approach provides a fast, differentiable surrogate that enables efficient forward evaluation and facilitates inverse analysis. The proposed neural operator treats the failure envelope as a ``continuous'' function, represented through irregularly sampled point sets. This enables learning from data with varying resolution and avoids reliance on fixed discretizations (i.e., is discretization-agnostic), which is essential when the data are heterogeneous \cite{bahmani2025resolution}.

In addition, to address the potential geometric inconsistencies observed in classical homogenization-based models, we incorporate a physics-informed regularization motivated by Drucker’s postulate. A curvature-based penalty is introduced to promote convexity of the learned failure envelopes, ensuring physically admissible responses while retaining fidelity to the underlying micromechanical model. 
The differential operators required for this regularization are computed using a finite difference (FD) scheme, in contrast to the more common use of automatic differentiation (AD). We empirically demonstrate that the FD approach provides faster training and inference.

The differentiable structure of the surrogate further enables the formulation of an inverse problem, where micromechanical configurations are identified to match a target failure envelope. Unlike traditional calibration approaches, this framework allows systematic and gradient-based optimization. 

To reduce the cost of data generation, we introduce an adaptive sampling (active learning) strategy that selectively queries the high-fidelity micromechanical model based on measures of uncertainty and novelty. This is particularly important in applications where data acquisition is expensive, either due to the high computational cost of simulations or reliance on costly experiments. This becomes even more critical in high-dimensional settings, where random or grid-based sampling leads to prohibitively large datasets. Moreover, increasing the dataset size alone does not necessarily improve the predictive performance of data-driven surrogate models; instead, informed and targeted sampling can significantly enhance sample efficiency and generalization \cite{liu2018survey,settles2009active}.

\section*{Related literature}

\textbf{Operator learning:} Operator learning has emerged as a powerful framework for approximating mappings between function spaces, with representative examples including Deep Operator Networks (DeepONet) \cite{lu2021learning}, the Fourier Neural Operator (FNO) \cite{li2020fourier}, and several subsequent variants tailored to partial differential equations and scientific computing \cite{seidman2022nomad,kovachki2023neural,wang2021learning,karumuri2020simulator,tripura2022wavelet,bahmani2025resolution,bahmani2025neural}.
More recently, operator-learning methods have also been extended to inverse problems, including neural inverse operators and related latent or physics-informed formulations for partial differential equation (PDE)-constrained inversion \cite{molinaro2023neural,wang2024latent}.
These methods have shown strong performance in learning forward or inverse maps in complex parametric systems. In contrast to standard PDE-solution settings, the present work focuses on learning macroscopic failure envelopes as continuous geometric objects generated by a micromechanics-based model, while also using the learned surrogate for inverse identification of micromechanical parameters.

\textbf{Scientific machine learning in solid and granular mechanics:}
Recent developments in scientific machine learning have enabled data-driven surrogates and reduced-order models for multiscale solid mechanics, constitutive modeling, and inverse problems \cite{karniadakis2021physics,bhaduri2022stress,wang2018multiscale}. These efforts include neural-network constitutive models \cite{fuhg2025review}, physics-informed surrogates \cite{haghighat2021physics}, and operator-learning approaches for parameterized boundary value problems \cite{yamazaki2025finite}. 
Related efforts include neural network surrogate models for yield surfaces in crystalline and metal-based materials \cite{nascimento2023machine,heidenreich2023modeling}.
For granular and quasi-brittle materials, however, learning frameworks that directly connect micromechanical configurations to macroscopic failure envelopes remain relatively limited \cite{qu2021towards,qu2023deep}. The present work contributes in this direction by constructing a differentiable surrogate for a homogenization-based granular micromechanics model.

\textbf{Differentiation schemes in physics-informed machine learning:}
In scientific machine learning, physics-informed approaches commonly leverage AD to construct PDE residuals \cite{karniadakis2021physics}.
As an alternative, FD schemes have been integrated to evaluate physics-informed losses in both multilayer perceptron (MLP) \cite{chiu2022can,lim2022physics} and neural operator architectures~\cite{liu2025physics}.
Theoretical and empirical comparisons between FD and AD suggest that AD can yield faster training convergence in the neural PDE solver setting~\cite{chen2025automatic}, while FD can be preferable in wall-clock time for MLP-based formulations~\cite{jiang2023applications}.
Separately, efficient batching of the physics loss has been achieved through separable architectures that combine per-axis sub-networks with forward-mode AD~\cite{cho2023separable}.
In the present work, we adopt FD for the curvature-based regularization and provide a systematic comparison with AD in terms of both forward evaluation time and training dynamics within a DeepONet architecture.

\textbf{Active learning in scientific machine learning:}
Active learning methods can be broadly categorized by how uncertainty is estimated and how queries are selected \cite{settles2009active}. Uncertainty estimation approaches range from Bayesian methods such as Gaussian processes and Monte Carlo dropout \cite{gal2016dropout}, to ensemble-based strategies that approximate epistemic uncertainty through model disagreement \cite{lakshminarayanan2017simple,krogh1994neural,seung1992query}. Query selection strategies range from purely exploitative methods targeting regions of high uncertainty, to exploratory methods promoting input-space coverage \cite{sener2017active}, to hybrid strategies balancing both \cite{kirsch2019batchbald,ash2019deep}. In scientific machine learning, active learning and adaptive sampling have been used to reduce the cost of data generation when labels are obtained from expensive simulations or experiments \cite{musekamp2024active,wu2023comprehensive}. 
In the context of operator learning, recent works include pool-based active learning benchmarks across nonlinear systems \cite{musekamp2024active} and multi-resolution operator learning with Fourier Neural Operators \cite{li2024multi}.
In this work, we adopt an ensemble-based approach due to its favorable scalability and ease of implementation compared to fully Bayesian alternatives. We employ a hybrid acquisition strategy that combines ensemble-based uncertainty in function space with a novelty measure relative to previously sampled configurations.

\subsection*{Paper Organization}
% \alert{Jinkyo complete this}

The remainder of the paper is organized as follows.
In Section~\ref{sec:formulation}, we present the micromechanical model and the formulation of the discretization-agnostic neural operator, as well as the associated curvature-based regularization, inverse identification and adaptive sampling strategies.
In Section~\ref{sec:results}, we compare the differentiation methods for the proposed penalty and the surrogate model, examine the effect of the proposed penalty, and present the results for inverse identification and adaptive sampling.
In Section~\ref{sec:conclusion}, we summarize the main findings.

\section{Formulation}\label{sec:formulation}
In this section, we first present, in Section \ref{sec:GMA}, the micromechanics-based model of granular materials that serves as the high-fidelity generator of failure envelopes. In Section \ref{sec:NO}, we introduce the operator learning framework for constructing a surrogate that maps microstructure configurations directly to failure envelopes, along with its inverse formulation for identifying microstructure configurations corresponding to a target response. Finally, in Section \ref{sec:AL}, we present an active learning framework that adaptively selects the most informative microstructures to be evaluated by the microscale solver, with the goal of reducing the amount of labeled data required for training.

\subsection{A Micromechanics-Based Model of Granular Materials}
\label{sec:GMA}

In this section, we briefly elaborate the general framework of the GMA, used to derive the behavior of granular materials from the collective interactions among grains. Within GMA, the material is envisioned as a collection of grains interacting with their neighbors through various inter-granular mechanisms. The modeling process involves three main steps: (i) a macro-micro kinematic identification step, whereby the grain-scale kinematics are derived from the macroscopic strain measures, (ii) a microscopic constitutive analysis, whereby the energetically conjugate forces developed at grain interactions are formulated, and (iii) a micro-macro upscaling process based on the Hill-Mandel micro-heterogeneity condition \cite{Hill1963}.

\begin{figure}[htb]
	\centering
	\includegraphics[scale=0.5]{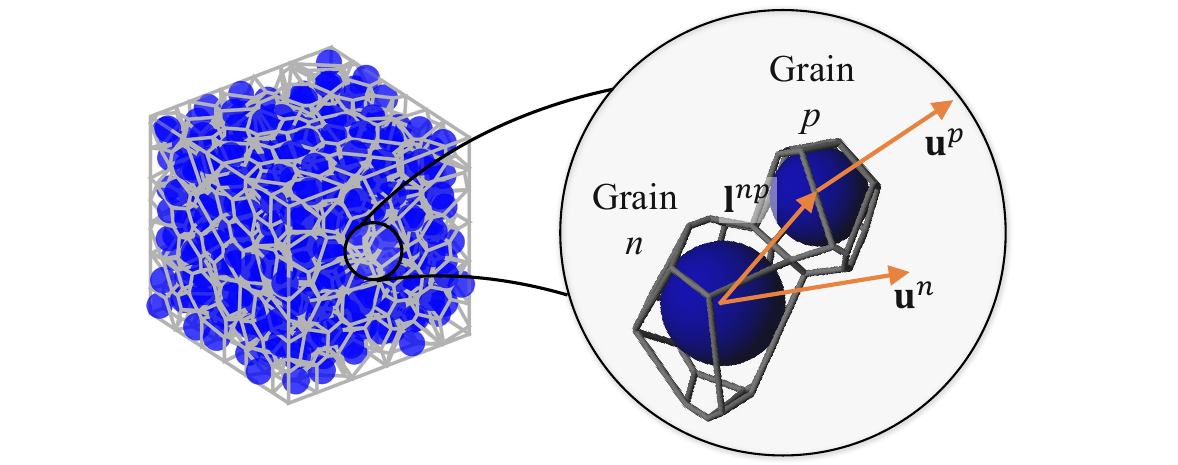}
	\caption{A tessellated granular system. Two neighboring grains and their displacements are magnified}
	\label{GMA-np}
\end{figure}

Within a given material point (see \cref{GMA-np}), consider two neighboring grains, $n$ and $p$, with centroidal positions $\mathbf{x}^n$ and $\mathbf{x}^p$, respectively. In a first-gradient analysis, the displacement of grain \textit{p}, denoted $\mathbf{u}^p$, can be approximated using a first-order Taylor expansion of the displacement field evaluated at grain \textit{n}, i.e., $\mathbf{u}^n$, as
$$
u_i^p \approx u_i^n + u_{i,j}^n \left(x_j^p - x_j^n\right).
$$
% where $u_i^n$ represents the Cartesian components of the displacement field evaluated at the centroid of grain $n$.
% at in a material point whose displacement field is given as $\mathbf{u} = (u_1, u_2, u_3)$. 
% where $x_i^p$ and $x_i^n$ are the centroidal positions of the two grains $n$ and $p$.
Throughout the paper, Einstein's summation convention over repeated indices is implied unless explicitly mentioned otherwise. Since the antisymmetric part of the displacement gradient corresponds to the rigid body rotation of the material point and, therefore, will not contribute to deformation energy, the relative displacement between the two grains can be found by replacing the displacement gradient with the strain tensor, yielding
\begin{equation}
	\delta_i^{np}=u_i^p-u_i^n \approx \epsilon_{ij}^n\left(x_j^p-x_j^n\right)=\epsilon_{ij}^nl_j^{np},
	\label{eq:di}
\end{equation}
where $l_j^{np}$ and $\epsilon_{ij}^n$ denote the Cartesian components of the branch vector joining the centroids of the two grains and the strain tensor at grain $n$.

% We note here that this relationship is valid when the material point's deformation is fully described using the first gradient of displacement, and higher gradients and grain rotations are neglected \cite{poorsolhjouy2019granular}.
Following the kinematic hypothesis, the strain tensor in \cref{eq:di} can be assumed to be constant within the material point. This assumption, also known as known as the Voigt assumption \cite{voigt1910lehrbuch}, implies that displacement fluctuations vanish over the material point and is therefore known to produce an upper bound for the stiffness of the material \cite{geers2017homogenization,dalbosco2021multiscale,dalbosco2024multiscale}. This assumption has been widely used for multiscale analysis of various materials, including granular \cite{Cha1988,Cha1996,liao2000,poorsolhjouy2025rock}, polymeric \cite{misra2013micromechanical}, and biological \cite{dalbosco2021multiscale,dalbosco2024multiscale} materials. Implemeting the kinematic hypothesis into \cref{eq:di}, the Cartesian components of the relative displacement $\delta_i^\alpha$ of grain-pair contact $\alpha$ is formulated as

\begin{equation}
	\delta_i^{\alpha} = \epsilon_{ij} l_j^\alpha.
	\label{eq:dialpha}
\end{equation}

To facilitate defining the grain-pair interactions in a physically meaningful manner, in each grain interaction, the displacement vector is written in terms of the following normal and tangential components
\begin{equation}
    \delta_\text{n}=\delta_i n_i = \epsilon_{ij}l_j n_i, \quad
    \delta_\text{w}=\sqrt{\left(\delta_i s_i\right)^2 + \left(\delta_i t_i\right)^2},
	\label{eq:dndw}
\end{equation}
where $n_i$, $s_i$ and $t_i$ represent the components of normal and two tangential unit base vectors constructed at the grain interaction. We note here that the variables $\delta_\text{n}$ and $\delta_\text{w}$ are \textit{scalar} values representing the inter-granular  displacement in the local coordinate directions. Therefore, throughout the paper the subscripts $\text{n}$ and $\text{w}$ \textit{do not} follow Einstein's summation convention.

We now endow each grain interaction, $\alpha$, with a deformation energy density, $W^\alpha$, such that the work-conjugate forces developed at the contact are derived from the partial derivatives of the energy potential with respect to the local displacements:
\begin{equation}
f_{\text{n}}^\alpha = \frac{\partial W^\alpha}{\partial \delta^\alpha_{\text{n}}}, \quad f_{\text{w}}^\alpha = \frac{\partial W^\alpha}{\partial \delta^\alpha_{\text{w}}}.
\label{eq:f-conj}
\end{equation}
Further, following from \cref{eq:f-conj}, the grain scale tangent stiffness coefficients in the normal and shear directions are defined as
\begin{equation}
k_{\text{n}}^\alpha = \frac{\partial f_{\text{n}}^\alpha}{\partial \delta^\alpha_{\text{n}}},
\quad
k_{\text{w}}^\alpha = \frac{\partial f_{\text{w}}^\alpha}{\partial \delta^\alpha_{\text{w}}}.
\label{eq:knkw}
\end{equation}

Finally, to derive the macroscopic stress tensor of the material point based on the behavior of grain-pair interactions, we employ the Hill-Mandel micro-heterogeneity condition \cite{Hill1963,miehe2002computational}, which implies that the incremental macroscopic strain energy density is equal to the volume average of the incremental energies of all grain interactions in the material point. Defining Cauchy stress as the work conjugate to the strain tensor, and substituting \cref{eq:dialpha,eq:f-conj} into the Hill-Mandel condition yields
\begin{equation}
    \sigma_{ij} = \frac{\partial W}{\partial \epsilon_{ij}} = \frac{1}{V} \sum_{\alpha=1}^ {N_\text{c}} \frac{\partial W^\alpha}{\partial \delta_k^\alpha} \frac{\partial \delta_k^\alpha}{\partial \epsilon_{ij}}  = \frac{1}{V} \sum_{\alpha=1}^ {N_\text{c}} f_i^\alpha l _j^\alpha,
    \label{eq:filj}
\end{equation}
where $N_\text{c}$ denotes the total number of grain-pair contacts in the material point.
Further, by defining the macroscopic stiffness tensor as the gradient of Cauchy stress tensor with respect to the strain tensor, and utilizing the kinematic assumption, \cref{eq:dialpha}, the components of the tangent stiffness tensor are derived as
\begin{equation}
    C_{ijkl} = \frac{\partial \sigma_{ij}}{\partial \epsilon_{kl}} = \frac{1}{V} \sum_{\alpha=1}^ {N_\text{c}} \frac{\partial f_i^\alpha}{\partial \delta_p^\alpha} \frac{\partial \delta_p^\alpha}{\partial \epsilon_{kl}} l _j^\alpha = \frac{1}{V} \sum_{\alpha=1}^ {N_\text{c}} k_{ik}^\alpha l _j^\alpha l _l^\alpha,
    \label{eq:Cijkl}
\end{equation}
where $k_{ik}^\alpha = \partial f_i^\alpha / \partial \delta_k^\alpha$ are the Cartesian components of the inter-granular tangent stiffness. Following tensor transformation laws, these can be defined in terms of the normal and shear stiffness coefficients, \cref{eq:knkw}, as
\begin{equation}
    k_{ik}^\alpha = k_\text{n}^\alpha n_i^\alpha n_k^\alpha + k_\text{w}^\alpha \left( s_i^\alpha s_k^\alpha +  t_i^\alpha t_k^\alpha \right).
    \label{kik}
\end{equation}

The relationship given in \cref{eq:Cijkl} yields the stiffness tensor for a grain packing composed of ${N_\text{c}}$ inter-granular contacts and relies on the explicit realization of the granular microstructure, involving the inter-granular stiffness coefficients and contact geometry. However, if the material point contains a sufficiently large number of particles, the summation over all contacts in \cref{eq:Cijkl} can be transformed into an integration over all orientations. 
For an isotropic granular material, where the geometric and mechanical properties of the particle contacts are distributed uniformly in all orientations, the macroscopic tangent can be rewritten in integral form as

\begin{equation}
C_{ijkl}=\frac{l^2 \rho_\text{c} }{4\pi} \int_{\theta=0}^\pi \int_{\phi=0}^{2\pi} k_{ik} n_j n_l \sin\theta \mathrm{d}\theta \mathrm{d}\phi,
\label{eq:Cijkl-int}
\end{equation}
with $l$, $\rho_\text{c}$, and $k_{ik}$ denoting the average grain size, density of grain interactions ($\rho_\text{c} \equiv N_\text{c}/V$), and the average inter-granular stiffness \cite{misra2016granular,pirmoradi2024anisotropic}.

The derivation of the tangent operator throughout the loading process allows us to define failure by tracking the macroscopic stiffness, rather than as a prescribed failure condition. Specifically, failure is defined as the loss of material stability, characterized by the singularity of the macroscopic tangent operator $\mathbb{C}$, derived in Eq. \eqref{eq:Cijkl-int} \cite{willam2002constitutive, rizzi1996failure,nicot2011diffuse}.  Since each contact direction undergoes its own loading history, the stress state at which $\mathbb{C}$ loses its positive-definiteness is dictated by the specific loading path followed \cite{poorsolhjouy2017effect, poorsolhjouy2025rock}, which allows us to predict failure as a path-function of the stress tensor.

\subsection{Neural Operator}
\label{sec:NO}

In this section, we present the neural operator as a differentiable forward surrogate that approximates the response of the GMA solver. This surrogate enables efficient evaluation of failure envelopes and provides the foundation for downstream tasks. We then introduce a formulation for inverse identification, where the differentiable structure of the neural operator is leveraged to recover microstructure configurations corresponding to target failure responses.
Finally, we present an adaptive sampling strategy that is integrated with the training process to generate informative data on the fly, improving data efficiency and enhancing the performance of the neural operator.

\subsubsection{Forward Surrogate Model: Microstructure-to-Failure-Envelope Mapping}
\label{sec:forward-oper-net}

In this section, we introduce a neural operator framework that serves as a differentiable surrogate, mapping microstructure configurations to their corresponding failure envelopes without explicit microstructure simulations.

A microstructure configuration is defined by a vector $\boldsymbol{\xi} \in \mathbb{R}^{p}$, where $p$ is the number of parameters in the micromechanics-based model.  We assume the existence of an underlying operator
\[
\mathcal{F} : \mathbb{R}^{p} \rightarrow \mathcal{V},
\]
where $\mathcal{V}$ is a function space describing the failure envelope. For a given microstructure $\boldsymbol{\xi}$, the output $\mathcal{F}(\boldsymbol{\xi})$ represents a continuous failure envelope.

Our goal is to approximate this operator using a parameterized neural operator. To obtain a discretization agnostic representation, we adopt an implicit formulation and define $\mathcal{F}_{\boldsymbol{\theta}}(\boldsymbol{\xi}, \boldsymbol{t}) \in \mathbb{R}^{q},$
where $\boldsymbol{t} \in \mathbb{R}^{l}$ is an auxiliary coordinate that continuously parameterizes the failure envelope. For a fixed $\boldsymbol{\xi}$, the collection of outputs $\{ \mathcal{F}_{\boldsymbol{\theta}}(\boldsymbol{\xi}, \boldsymbol{t}) \}$ defines an implicit representation of the failure surface. The parameters $\boldsymbol{\theta}$ are learned so that this surface approximates $\mathcal{F}(\boldsymbol{\xi})$.

In practice, the failure envelope is observed through a finite set of sampled stress states. We denote this discretization by
\begin{equation}
    \mathcal{Y} =
    \left\{
        \boldsymbol{\sigma}^{(i)} \in \mathbb{R}^{q}
    \right\}_{i=1}^{N},
\end{equation}
where each $\boldsymbol{\sigma}^{(i)}$ lies on the failure envelope and corresponds to samples of $\mathcal{F}(\boldsymbol{\xi})$.
Note that $\mathcal{Y}$ is an unordered set, and the number of points $N$ may vary across different failure envelopes.

Given this problem setting, we assume that data are available, or can be queried from a black-box simulator or experimental setup, in the form
\[
\left\{ \boldsymbol{\xi}^{(m)}, \mathcal{Y}^{(m)} \right\}_{m=1}^{M},
\]
where each $\mathcal{Y}^{(m)}$ is an unordered set of $N_m$ points.

The neural operator parameters are trained by minimizing the expected relative mean squared error (MSE) between the predicted and observed point sets:
\begin{equation}\label{eq:loss_rel_mse}
\mathcal{L}_{\text{relMSE}}(\boldsymbol{\theta})
=
\mathbb{E}_{(\boldsymbol{\xi},\mathcal{Y})}
\left[
\mathrm{relMSE}\!\left(
\widehat{\mathcal{Y}}_{\boldsymbol{\theta}}(\boldsymbol{\xi}),
\mathcal{Y}
\right)
\right],
\end{equation}
where the predicted point set is defined as
\begin{equation}
\widehat{\mathcal{Y}}_{\boldsymbol{\theta}}(\boldsymbol{\xi})
=
\left\{
\mathcal{F}_{\boldsymbol{\theta}}(\boldsymbol{\xi}, \boldsymbol{t}^{(j)})
\right\}_{j=1}^{N_{{t}}},
\end{equation}
with $\{ \boldsymbol{t}^{(j)} \}_{j=1}^{N_{{t}}}$ denoting a fixed set of auxiliary (latent) coordinates used to query the neural operator.
The relative MSE is defined as
\begin{equation}\label{eq:rel_MSE}
    \mathrm{relMSE}
    % (\boldsymbol{\xi})
    \left(
    \widehat{\mathcal{Y}}_{\boldsymbol{\theta}}(\boldsymbol{\xi}),
    \mathcal{Y}
    \right)
    =
    \frac{
    \frac{1}{N}
    \sum_{i=1}^{N}
    \left\|
    \mathcal{F}_{\boldsymbol{\theta}}
    (\boldsymbol{\xi},\boldsymbol{t}^{(i)})
    -
    % \hat{\mathcal{Y}}
    \boldsymbol{\sigma}^{(i)}
    % (\boldsymbol{t};\boldsymbol{\xi})
    \right
    \|^2
    }{
    \frac{1}{N}
    % \sum_{\boldsymbol{t}\in\mathcal{T}}
    \sum_{i=1}^{N}
    \left\|
    % \hat{\mathcal{Y}}(\boldsymbol{t};\boldsymbol{\xi})
    \boldsymbol{\sigma}^{(i)}
    \right\|^2
    +\varepsilon
    },
\end{equation}
where we choose $\varepsilon=10^{-12}$ for numerical stability.
Here, we use the $L_2$ norm, and $\mathcal{F}_{\boldsymbol{\theta}}(\boldsymbol{\xi},\boldsymbol{t}^{(i)})\in\widehat{\mathcal{Y}}_{\boldsymbol{\theta}}(\boldsymbol{\xi})$ represents the high-fidelity microscale simulator output corresponding to the query point $\boldsymbol{t}^{(i)}$ for the given parameter $\boldsymbol{\xi}$, that is, one component of the simulator response $\boldsymbol{\sigma}^{(i)}\in\mathcal{Y}$.

\begin{figure}[htbp]
    \centering
    \begin{subfigure}{0.68\textwidth}
        \centering
        \includegraphics[width=\textwidth]{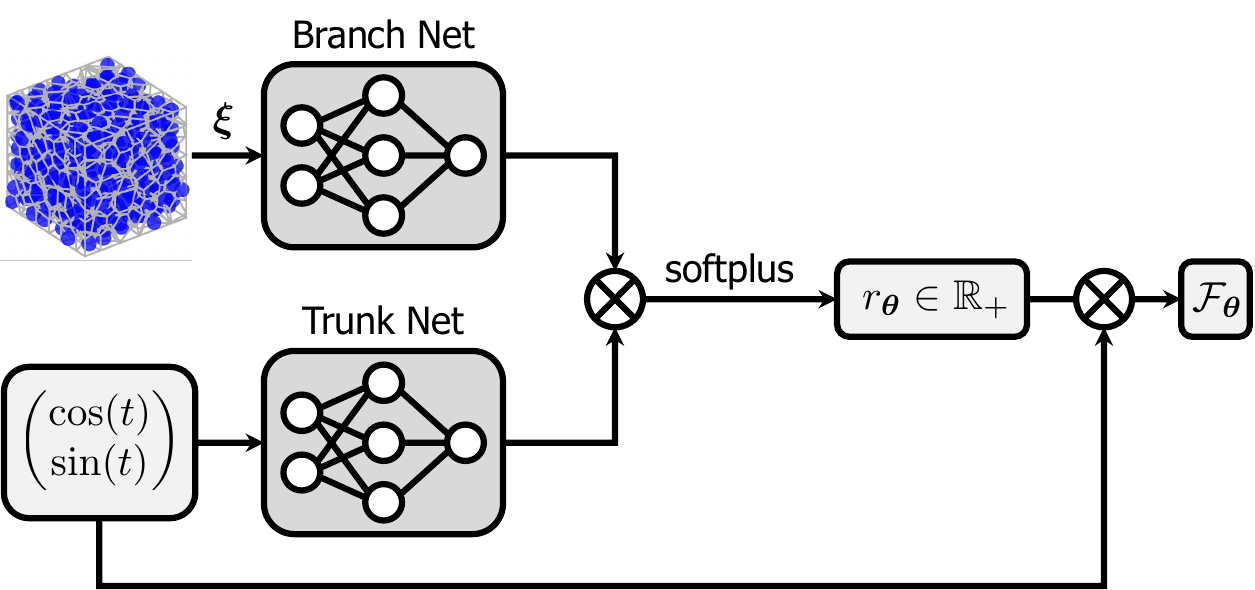}
    \end{subfigure}
    \hfill
    \caption{
    Schematic illustration of the architecture used to parametrize the surrogate models.
    }
    \label{fig:architecture}
\end{figure}

While the neural operator is formulated in terms of a general latent coordinate $\boldsymbol{t} \in \mathbb{R}^{l}$, we restrict attention here to failure envelopes represented as planar curves ($q=2$). In this case, the envelope admits a one-dimensional manifold structure and can be parameterized by a scalar coordinate $t \in \mathbb{R}$ (e.g., arc-length or an equivalent monotone parameterization).

We represent $\mathcal{F}_{\boldsymbol{\theta}}(\boldsymbol{\xi}, t)$ in polar form:

\begin{equation}\label{eq:parametrization}
\mathcal{F}_{\boldsymbol{\theta}}(\boldsymbol{\xi}, t)
=
\big[
r_{\boldsymbol{\theta}}(\boldsymbol{\xi}, \cos t, \sin t)\cos t,
r_{\boldsymbol{\theta}}(\boldsymbol{\xi}, \cos t, \sin t)\sin t
\big]^\top,
\end{equation}
where $r_{\boldsymbol{\theta}} : \mathbb{R}^{p+2} \to \mathbb{R}_+$ is modeled using a DeepONet architecture \cite{lu2021learning} with a softplus output to enforce positivity: 
\begin{equation}
    r_{\boldsymbol{\theta}}
    (\boldsymbol{\xi}, \cos(t),\sin(t))
    =
    \mathrm{Softplus}
    \Bigl(
    \sum_{i=1}^{N_{p}}
    \mathrm{br}_{\boldsymbol{\theta}}^{(i)}
    (\boldsymbol{\xi})\,
    \mathrm{tr}_{\boldsymbol{\theta}}^{(i)}
    (\cos(t),\sin(t))
    +b
    \Bigr),
\end{equation}
where $N_p$, $b$, $\mathrm{br}_{\boldsymbol{\theta}}^{(i)}$, and $\mathrm{tr}_{\boldsymbol{\theta}}^{(i)}$ denote the latent width (number of produced basis functions by trunk network), the bias (intercept) term, and the $i$-th components of the branch and trunk networks, respectively.
The branch network takes the microstructure configuration $\boldsymbol{\xi}$ as input, while the trunk network takes $(\cos t, \sin t)$ as input, ensuring $2\pi$-periodicity in $t$. The overall architecture is illustrated in \cref{fig:architecture}.

To enforce mechanical admissibility consistent with Drucker's postulate, we augment the training objective with a curvature-based regularization that promotes convexity of the learned failure envelope.
Under the parameterization in \cref{eq:parametrization}, convexity of the failure envelope can be assessed through the sign of its curvature.
The signed curvature at $t$ is given by
\begin{equation}\label{eq:curvature}
\kappa_{\boldsymbol{\theta}}(t;\boldsymbol{\xi})
=
\frac{
\det\!\left(
\partial_t \mathcal{F}_{\boldsymbol{\theta}}(\boldsymbol{\xi}, t),
\partial_{tt} \mathcal{F}_{\boldsymbol{\theta}}(\boldsymbol{\xi}, t)
\right)
}{
\left\|
\partial_t \mathcal{F}_{\boldsymbol{\theta}}(\boldsymbol{\xi}, t)
+\varepsilon
\right\|_2^3
},
\end{equation}
which corresponds to the signed curvature of the parametric curve 
% \alert{REF}
\cite{do2016differential}.
We choose $\varepsilon=10^{-12}$ for numerical stability.
In this work, the derivatives $\partial_{t}\mathcal{F}_{\boldsymbol{\theta}}$ and $\partial_{tt}\mathcal{F}_{\boldsymbol{\theta}}$ are approximated using central finite differences with periodic boundary conditions on the equispaced grid $\mathcal{T}$.

Based on this quantity, we define the curvature penalty
\begin{equation}\label{eq:curvature_penalty}
\mathcal{L}_{\kappa}(\boldsymbol{\theta})
=
\mathbb{E}_{\boldsymbol{\xi}}
\left[
\frac{1}{|\mathcal{T}|}
\sum_{t \in \mathcal{T}}
\max\!\left(
-\kappa_{\boldsymbol{\theta}}(t;\boldsymbol{\xi}),\, 0
\right)
\right],
\end{equation}
where $\mathcal{T}$ is a set of sampled parameter values used to evaluate the curvature.
The sign convention depends on the chosen orientation of the parameterization.
Here, we adopt the negative sign, since the parameterization in \cref{eq:parametrization} traverses the curve in the counterclockwise direction.

The overall training objective combines the relative MSE with the curvature regularization:
\begin{equation}\label{eq:loss_ftn}
\mathcal{L}(\boldsymbol{\theta})
=
\mathcal{L}_{\mathrm{relMSE}}(\boldsymbol{\theta})
+
\lambda \mathcal{L}_{\kappa}(\boldsymbol{\theta}),
\end{equation}
where $\lambda$ is a penalty parameter.
In this work, the curvature loss is defined using only the numerator of ~\cref{eq:curvature}, as the regularization only requires enforcing the sign of the curvature, not its normalized magnitude.

\section*{Efficient Batch-wise Differentiation}
In this section, we describe an efficient batch-wise evaluation of the first- and second-order derivatives required for computing the curvature loss function. This efficiency stems from the low-rank structure of the DeepONet backbone used here \cite{cho2023separable}.

For the curvature loss computation, the derivatives of the point coordinates in the stress space with respect to the angular coordinate $t$ are required, which in turn depend on the gradients of the parametrized radius $r_{\boldsymbol{\theta}}$. For notational convenience, we denote the preactivation of the {softplus} operator by $z$.

Since the trunk network only depends on $t$, its derivatives $\partial_t\mathrm{tr}_{\boldsymbol{\theta}}^{(i)}(t)$ and $\partial_{tt}\mathrm{tr}_{\boldsymbol{\theta}}^{(i)}(t)$ are independent of the batched parameters $\boldsymbol{\xi}$. Therefore, they can be evaluated once by differentiation through the trunk network alone, rather than recomputed for each parameter sample. Let $\{t_j\}_{j=1}^{N_t}$ denote the angular collocation points and $\{\boldsymbol{\xi}_s\}_{s=1}^{N_s}$ the parameter batch. The branch and trunk evaluations are collected into the matrices
\begin{equation}
    \mathrm{br}_{si}
    =\mathrm{br}_{\boldsymbol{\theta}}^{(i)}(\boldsymbol{\xi}_s),\quad
    \mathrm{tr}_{ji}
    =\mathrm{tr}_{\boldsymbol{\theta}}^{(i)}(t_j),\quad
    \mathrm{tr}^{\prime}_{ji}
    =\partial_t\mathrm{tr}_{\boldsymbol{\theta}}^{(i)}(t_j),\quad
    \mathrm{tr}^{\prime\prime}_{ji}
    =\partial_{tt}\mathrm{tr}_{\boldsymbol{\theta}}^{(i)}(t_j),
\end{equation}
of sizes $N_s\times N_p$ and $N_t\times N_p$, respectively.
The pre-activation field and its first two derivatives over the entire batch are then assembled through matrix products,
\begin{equation}
    z_{sj}
    =
    \sum_{i=1}^{N_p}
    \mathrm{br}_{si}\,\mathrm{tr}_{ji}+b,\quad
    \partial_t z_{sj}
    =
    \sum_{i=1}^{N_p}
    \mathrm{br}_{si}\,\mathrm{tr}^{\prime}_{ji},\quad
    \partial_{tt} z_{sj}
    =
    \sum_{i=1}^{N_p}
    \mathrm{br}_{si}\,\mathrm{tr}^{\prime\prime}_{ji}.
\end{equation}
The softplus chain rule and the trigonometric product rule are then applied element-wise over $(s,j)$ to recover $\partial_t r_{\boldsymbol{\theta}}$, $\partial_{tt} r_{\boldsymbol{\theta}}$, and the corresponding Cartesian derivatives.
Because the trunk differentiation is performed once and reused across the entire parameter batch, the cost of automatic differentiation is independent of $N_s$; the remaining assembly consists of matrix products and element-wise operations of cost $O(N_s N_t N_p)$.
The curvature at each $(s,j)$ is then evaluated as
\begin{equation}
    \kappa_{sj}
    =
    \frac{
        \partial_t x\,\partial_{tt} y
        -\partial_t y\,\partial_{tt} x
    }{
        \bigl(
        (\partial_t x)^2+(\partial_t y)^2
        +\varepsilon
        \bigr)^{3/2}
    },
\end{equation}
which is equivalent to \cref{eq:curvature}.
We choose $\varepsilon=10^{-12}$ for numerical stability.

\subsubsection{Inverse Identification: Failure Envelope to Microstructure}

In this section, we formulate the inverse identification problem of determining the microstructure configuration that realizes a prescribed target failure envelope. This problem is central to both material characterization and goal-oriented design.

Let $\mathcal{F}_{\boldsymbol{\theta}}(\boldsymbol{\xi}, \boldsymbol{t})$ denote the trained neural operator. For a given $\boldsymbol{\xi}$, the corresponding failure envelope is represented as a point cloud
\begin{equation}
\widehat{\mathcal{Y}}_{\boldsymbol{\theta}}(\boldsymbol{\xi})
=
\left\{
\mathcal{F}_{\boldsymbol{\theta}}(\boldsymbol{\xi}, \boldsymbol{t}^{(j)})
\right\}_{j=1}^{N_t},
\end{equation}
where $\{ \boldsymbol{t}^{(j)} \}_{j=1}^{N_t}$ are auxiliary coordinates used to query the operator.

Given a target failure envelope represented as an unordered point set
\[
\mathcal{Y} =
\left\{
\boldsymbol{\sigma}^{(i)}
\right\}_{i=1}^{N},
\]
the inverse problem is formulated as
\begin{equation}
\boldsymbol{\xi}^{\ast}
=
\arg\min_{\boldsymbol{\xi} \in \mathcal{X}}
\mathcal{J}(\boldsymbol{\xi}),
\end{equation}
where $\mathcal{X} \subset \mathbb{R}^{p}$ denotes the admissible parameter space. The mismatch functional is defined using the relative MSE \cref{eq:rel_MSE}.

In practice, physically admissible ranges for the microstructure parameters are typically known \textit{a priori}, defining a bounded feasible set $\mathcal{X}$. We map this set to the normalized domain $\mathcal{X} = [-1,1]^p$ for numerical stability and consistency with training.

The inverse problem is solved using a projected gradient-based method. At each iteration, the parameters are updated using ADAM \cite{kingma2014adam} and projected back onto the feasible set to enforce admissibility. Convergence is assessed using the projected gradient residual
\begin{equation}
\label{eq:gradient_residual}
\mathcal{R}
=
\left\|
\Pi_{[-1,1]^p}
\bigl(
% \tilde{\boldsymbol{\xi}}
\boldsymbol{\xi}
-
\boldsymbol{\nabla}_{
% \tilde{\boldsymbol{\xi}}
\boldsymbol{\xi}
}\mathcal{J}(
% \tilde{\boldsymbol{\xi}}
\boldsymbol{\xi}
)
\bigr)
-
% \tilde{\boldsymbol{\xi}}
\boldsymbol{\xi}
\right\|_2,
\end{equation}
where $\Pi_{[-1,1]^p}$ denotes the Euclidean projection onto the box $[-1,1]^p$.

To mitigate sensitivity to initialization and potential non-uniqueness, the optimization is performed independently from multiple initial guesses sampled from the training distribution.

\subsection{Active Learning Strategy}
\label{sec:AL}
In this section, we introduce an adaptive sampling strategy that jointly trains the neural operator and generates new samples within a mixed adaptive sequential sampling framework \cite{eason2014adaptive}.

At each iteration, we maintain a labeled dataset 
\[
\mathcal{D} = 
    \{ 
        (
        \boldsymbol{\xi}_i, 
        \mathcal{F}(\boldsymbol{\xi}_i)
        ) 
    \}_{i=1}^{N}
\]
% and a pool of unlabeled candidate microstructure configurations
and sample a set of unlabeled candidate \cite{crombecq2011efficient} microstructure configurations,
\[
\mathcal{P} = \{ \boldsymbol{\xi}_j \}_{j=1}^{N_{\text{can}}}, \quad \boldsymbol{\xi} \in \mathbb{R}^p.
\]
The goal is to select a subset of informative candidates from $\mathcal{P}$, query their labels (via microscale simulation or experiment), and augment the dataset $\mathcal{D}$.

To quantify informativeness, we employ an ensemble of $N_{\text{ens}}$ neural operators $\{ \mathcal{F}_{\boldsymbol{\theta}_k} \}_{k=1}^{N_{\text{ens}}}$ trained on the current dataset $\mathcal{D}$. The disagreement among ensemble predictions serves as a proxy for epistemic uncertainty. For a candidate microstructure $\boldsymbol{\xi} \in \mathcal{P}$, we define an uncertainty score $\mathcal{U}(\boldsymbol{\xi})$ based on the variability of the predicted failure envelopes across the ensemble \cite{krogh1994neural}.

We first define the ensemble mean prediction
\[
\bar{\mathcal{F}}(\boldsymbol{\xi})
=
\frac{1}{N_{\text{ens}}}
\sum_{k=1}^{N_{\text{ens}}}
\mathcal{F}_{\boldsymbol{\theta}_k}(\boldsymbol{\xi}).
\]

We then define the uncertainty score using the unbiased estimator of the ensemble normalized variance in the function space $\mathcal{V}$,
\[
\mathcal{U}(\boldsymbol{\xi})
=
\frac{1}{N_{\text{ens}} - 1}
\sum_{k=1}^{N_{\text{ens}}}
\left\|
\mathcal{F}_{\boldsymbol{\theta}_k}(\boldsymbol{\xi})
-
\bar{\mathcal{F}}(\boldsymbol{\xi})
\right\|_{\mathcal{V}}^2
\,
\big/
\left(
\|\bar{\mathcal{F}}(\boldsymbol{\xi})\|_{\mathcal{V}}^2 + \varepsilon
\right),
\]

where $\|\cdot\|_{\mathcal{V}}$ denotes a suitable norm on the failure envelope space (e.g., an $L^2$ norm over the envelope parameterization), and $\varepsilon > 0$ is a small constant introduced for numerical stability. This normalization renders the uncertainty score scale-invariant with respect to the magnitude of the predicted failure envelope.

To avoid redundant sampling and promote coverage of the design space, we additionally introduce a novelty measure that penalizes candidates that are close to previously labeled samples. For a candidate $\boldsymbol{\xi} \in \mathcal{P}$, we define the novelty score
\[
\mathcal{N}(\boldsymbol{\xi})
=
\min_{\boldsymbol{\xi}_i \in \mathcal{D}}
d\!\left(\boldsymbol{\xi}, \boldsymbol{\xi}_i\right),
\]
where $d(\cdot,\cdot)$ denotes a suitable distance metric.
In this work, $d$ can be defined either in the microstructure parameter space (e.g., Euclidean distance in $\mathbb{R}^p$) or in a learned latent representation.

The final acquisition score balances these two normalized components,
\begin{equation}\label{eq:acquisition_tot}
\mathcal{A}(\boldsymbol{\xi}) 
= 
\tau \, \tilde{\mathcal{U}}(\boldsymbol{\xi}) 
+ 
(1 - \tau)\, \tilde{\mathcal{N}}(\boldsymbol{\xi}),
\end{equation}
where $\tau \in [0,1]$ controls the exploration--exploitation trade-off 
% \alert{REF}
\cite{liu2018survey}. In this work, we use $\tau=0.5$.

% To ensure comparability between the two terms, we normalize each score over the candidate pool $\mathcal{P}$.
To ensure comparability between the two terms, we normalize each score over the candidate set $\mathcal{P}$.
Specifically, we define
\[
\tilde{\mathcal{U}}(\boldsymbol{\xi})
=
\mathcal{U}(\boldsymbol{\xi})
\big/
\max_{\boldsymbol{\xi}' \in \mathcal{P}}
\mathcal{U}(\boldsymbol{\xi}'),
\qquad
\tilde{\mathcal{N}}(\boldsymbol{\xi})
=
\mathcal{N}(\boldsymbol{\xi})
\big/
\max_{\boldsymbol{\xi}' \in \mathcal{P}}
\mathcal{N}(\boldsymbol{\xi}').
\]

Among all candidates in $\mathcal{P}$, those with the top-$\mathrm{K}$ acquisition scores $\mathcal{A}(\boldsymbol{\xi})$ are selected, evaluated using a high-fidelity microscale model (or experiment), removed from $\mathcal{P}$, and appended to the dataset $\mathcal{D}$.
The neural operator is then retrained, and the process is repeated.

For the retraining procedure in the adaptive sampling iterations, we use the same relative MSE defined in Eq.~\ref{eq:rel_MSE}, averaged over both the parameter batch and the ensemble members. Specifically, the training loss is defined as
\begin{equation}\label{eq:loss_ensemble}
    \mathcal{L}
    =
    \mathbb{E}_{
    (\boldsymbol{\xi},\mathcal{Y})
    }
    \left[
        \frac{1}{N_{\mathrm{ens}}}
        \sum_{e=1}^{N_{\mathrm{ens}}}
        \mathrm{relMSE}
        \bigl(
            \widehat{\mathcal{Y}}_{\boldsymbol{\theta}}
            (\boldsymbol{\xi}),
            \mathcal{Y}
            % \boldsymbol{f}_{\boldsymbol{\theta}}^{(e)}
        \bigr)
    \right],
\end{equation}

The adaptive sequential sampling algorithm is illustrated in Algorithm~\ref{alg:active_learning}.

\begin{algorithm}[htbp]
\caption{Adaptive Sampling with Budget Constraint}
\label{alg:active_learning}
\begin{algorithmic}[1]
\State \textbf{Given:} microscale oracle $\mathcal{F}$, initial dataset $\mathcal{D}_0$, parameter space $\mathcal{X}$, ensemble size $N_{\text{ens}}$, sample size $N_{\text{can}}$, batch size $\mathrm{K}$, and budget $N_{\text{bud}}$.
\State Initialize ensemble $\{ \mathcal{F}_{\boldsymbol{\theta}_k} \}_{k=1}^{N_{\text{ens}}}$
\State $\mathcal{D} \gets \mathcal{D}_0$
\For{$i = 1,\dots,N_{\text{bud}}$}
    \State Train ensemble $\{ \mathcal{F}_{\boldsymbol{\theta}_k} \}_{k=1}^{N_{\text{ens}}}$ on $\mathcal{D}$
    \State Sample $\mathcal{P}\subset\mathcal{X}$ such that $|\mathcal{P}|=N_{\text{can}}$
    \State Compute acquisition scores $\mathcal{A}(\boldsymbol{\xi})$ for all $\boldsymbol{\xi} \in \mathcal{P}$
    \State Select
    \[
    \mathcal{P}_{\text{new}}
    =
    \operatorname{TopK}_{\boldsymbol{\xi} \in \mathcal{P}}
    \mathcal{A}(\boldsymbol{\xi})
    \]
    \State Evaluate $\mathcal{F}(\boldsymbol{\xi})$ for all $\boldsymbol{\xi} \in \mathcal{P}_{\text{new}}$
    \State Update dataset
    \[
    \mathcal{D}
    \gets
    \mathcal{D}
    \cup
    \{ (\boldsymbol{\xi}, \mathcal{F}(\boldsymbol{\xi})) \mid \boldsymbol{\xi} \in \mathcal{P}_{\text{new}} \}
    \]
    % \State Update pool $\mathcal{P} \gets \mathcal{P} \setminus \mathcal{P}_{\text{new}}$
\EndFor
% \State \textbf{Return} trained ensemble $\{ \mathcal{F}_{\boldsymbol{\theta}_k} \}_{k=1}^{N_{\text{ens}}}$
\State \textbf{Return} trained ensemble $\{ \mathcal{F}_{\boldsymbol{\theta}_k} \}_{k=1}^{N_{\text{ens}}}$, dataset $\mathcal{D}$
\end{algorithmic}
\end{algorithm}

\section{Results}\label{sec:results}

In this section, we evaluate the proposed framework and examine the effect of its key components on data generated from the microscopic model. The data generation procedure is described in \cref{sec:data}. A systematic comparison of the FD and AD schemes is presented in \cref{sec:results:FD}. The effect of the proposed curvature penalty is investigated in \cref{sec:results:surrogate}. Inverse identification is demonstrated in \cref{sec:results:inverse}. Finally, the efficiency of the adaptive sampling strategy is evaluated in \cref{sec:results:adaptive}. All experiments are performed in double precision.

\subsection{Data Generation and micromechanics simulator}\label{sec:data}

The general framework for GMA developed in Section \ref{sec:GMA} is used here to predict the failure envelope of generic cohesive quasi-brittle granular materials.
For this, we focus on cemented granular materials and rocks whose macroscopic response is characterized by a finite tensile strength and a higher compressive strength \citep{consoli2010parameters,diambra2018modelling}. 
To retain the irreversible nature of the damage process, resulting in path-dependent material response, in all different mechanisms, the inter-granular forces are defined using the following secant stiffness relations \cite{simo1998computational,ortiz1985constitutive}
\begin{equation}
    f_{\text{n}} = 
    \begin{dcases} 
        A(1 - \omega_{\text{nt}})\delta_{\text{n}} & \text{if } \delta_{\text{n}} \geq 0 \text{ (tension)} \\
        A(1 - \omega_{\text{nc}})\delta_{\text{n}} & \text{if } \delta_{\text{n}} < 0 \text{ (compression)} 
    \end{dcases}, \quad
    f_{\text{w}} = C(1 - \omega_{\text{w}})\delta_{\text{w}},
    \label{eq:fnfw-damage}
\end{equation}
with $A$ and $C$ denoting the initial stiffness of the intact contact in normal and tangential directions, respectively. Further $\omega_{\text{nt}}$, $\omega_{\text{nc}}$, and $\omega_{\text{w}}$ denote the damage variables in normal tension, normal compression, and shear, respectively. To ensure that in all mechanisms, the damage retains its irreversible nature, all damage variables are formulated as functions of their corresponding deformation history variables that track the maximum displacements experienced in different kinematic components.
Explicitly, $\kappa_{\text{nt}}$ and $\kappa_{\text{nc}}$ are introduced as the maximum tensile normal displacement and the maximum compressive normal displacement experienced by each contact. Similarly, $\kappa_{\text{w}}$ denotes the maximum tangential displacement experienced by each contact during the entire loading history.

The inter-granular constitutive relationships at the contact level are modeled using exponential force–displacement laws \cite{poorsolhjouy2017effect,mourlas2023large,pardoen2020accounting,Ma2016AGU,Ma2017AGU,wang2024failure}, as given below, which are commonly adopted to capture failure and post-peak softening behavior in cohesive granular materials:
\begin{equation}
    \omega_{\text{nt}} = 1 - \exp\left(-\frac{\kappa_{\text{nt}}}{B_{\text{t}}}\right), \quad
    \omega_{\text{nc}} = 1 - \exp\left(-\frac{\kappa_{\text{nc}}}{B_{\text{c}}}\right), \quad
    \omega_{\text{w}} = 1 - \exp\left(-\frac{\kappa_{\text{w}}}{D}\right),
    \label{damagevars}
\end{equation}
where the parameters $B_\text{t}$ and $B_\text{c}$ denote the tensile and compressive normal displacements at peak normal tensile and compressive force, respectively. Similarly, $D$ denotes the tangential displacement at which the tangential force reaches its maximum.
Since damage remains unchanged during unloading and reloading phases, unloading follows the elastic branch along a linear secant path connecting the onset of unloading to the origin.

\begin{figure}[htbp]
    \centering
    \begin{subfigure}{0.3\textwidth}
        \centering
        \includegraphics[width=\textwidth]{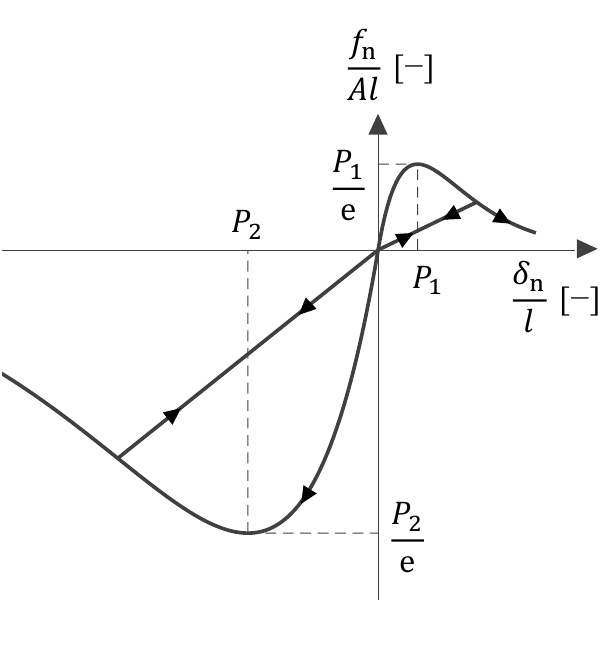}
        \caption{}\label{fig:fn}
    \end{subfigure}
    \hspace{1cm}
    \begin{subfigure}{0.3\textwidth}
        \centering
        \includegraphics[width=\textwidth]{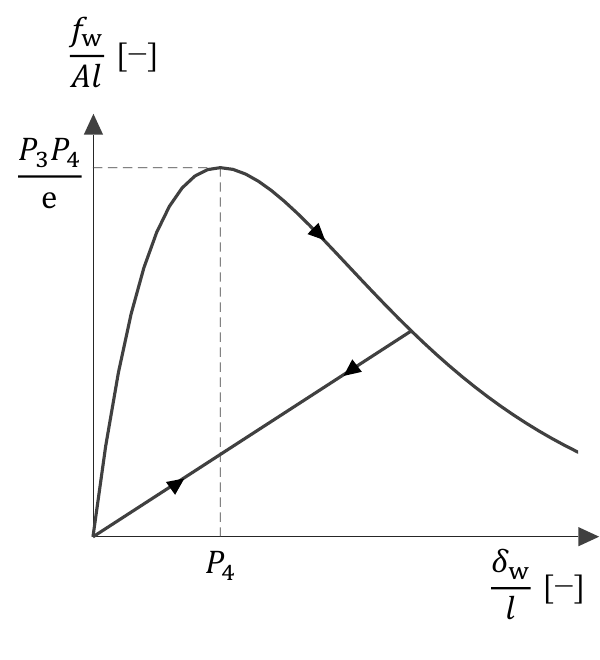}
        \caption{}\label{fig:fw}
    \end{subfigure}
    \hfill
    
    \caption{
    Dimensionless representation of the inter-granular constitutive relationships including unloading/reloading branches: (\subref{fig:fn}) normalized normal force-displacement relationship and (\subref{fig:fw}) normalized shear force-displacement relationship. The axes are non-dimensionalized using the initial normal stiffness $A$ and average grain size $l$. The characteristic peak forces and their corresponding displacements are explicitly identified by the vector of dimensionless parameters $\boldsymbol{\xi} = (P_1, P_2, P_3, P_4)$.
    }
    \label{fig:fd}
\end{figure}

To explore the diverse landscape of failure envelopes of granular materials, while maintaining a consistent physical scale, we identify a set of dimensionless parameters that govern the \textit{shape} and \textit{topology} of the failure surface rather than its absolute magnitude.
We treat the initial normal stiffness $A$, the average grain size $l$, and the contact density $\rho_{\text{c}}$ as fixed scaling constants. These parameters primarily act as linear multipliers for the stress tensor and do not alter the fundamental geometry of the failure envelope. The variation of the remaining constitutive parameters (those defining the shape of the failure envelope), i.e. the characteristic lengths ($B_{\text{t}}, B_{\text{c}}, D$) and the initial shear stiffness ($C$), are mapped to a vector of dimensionless parameters $\boldsymbol{\xi}=(P_1,P_2,P_3,P_4)$, defined as:
\begin{equation}
    P_1 = \frac{B_{\text{t}}}{l},
    \quad
    P_2 = \frac{B_{\text{c}}}{l},
    \quad
    P_3 = \frac{C}{A},
    \quad
    P_4 = \frac{D}{l}.
    \label{P1-4}
\end{equation}
The dimensionless parameters $P_1$ and $P_2$ represent the \textit{normalized ductility} in the normal direction, or characteristic softening lengths in tension and compression, respectively; $P_3$ defines the ratio between the initial shear stiffness to normal stiffness; and $P_4$ represents the \textit{normalized ductility} in the shear direction. \Cref{fig:fd} illustrates the resulting normalized inter-granular force-displacement relationships based on this parametrization. 

For the training results reported in \cref{sec:results:FD,sec:results:surrogate,sec:results:inverse}, the dataset is generated by sampling the four-dimensional micromechanical parameter space on a structured grid. Specifically, $10$ equispaced values are selected for each parameter:
$P_1 \in [0.01,\,0.19]$, 
$P_2 \in [-0.37,\,-0.01]$, 
$P_3 \in [0.20,\,0.875]$, and 
$P_4 \in [0.01,\,0.145]$,
resulting in a total of $10^4$ parameter combinations. This range allows the model to exhibit a rich variety of biaxial failure envelopes, spanning from the highly asymmetric ``teardrop'' or ``bullet-shaped'' surfaces characteristic of brittle rocks and over-consolidated clays \cite{hoek1980empirical, lade1975elastoplastic} to the more symmetric, cohesive-frictional envelopes found in cemented sands \cite{consoli2010parameters, diambra2018modelling}. Furthermore, the flexibility of the underlying micromechanical law enables the framework to capture a spectrum of failure behaviors: from the nearly linear envelopes associated with the classical Mohr-Coulomb criterion \cite{labuz2014mohr} to the highly non-linear, curved envelopes typical of the Drucker-Prager \cite{drucker1952soil} or Hoek-Brown \cite{hoek2019hoek} models.

For each parameter combination, the failure envelope in the $\sigma_{11} - \sigma_{22}$ stress space is numerically identified by loading the material along distinct radial (proportional) loading paths. To generate these envelopes, the model is subjected to incremental stress-controlled loading, where the unknown strain increments are determined via an explicit Forward Euler scheme. A detailed description of this numerical implementation is provided in Appendix \ref{apdx:ForwardEuler}. Following the stability criterion discussed in Section \ref{sec:GMA}, failure is identified as the point where the macroscopic tangent stiffness tensor $\mathbb{C}$ becomes singular, indicating the onset of material instability or loss of load-bearing capacity.

All input parameters are normalized to $[-1,1]$ via min--max scaling and concatenated with $\cos(t)$ and $\sin(t)$, where $t\in[0,2\pi)$, to ensure that the surrogate produces closed paths.
The latent coordinate set $\mathcal{T}$ is chosen as an equispaced grid with $144$ points over $[0,2\pi)$ for both training and evaluation, unless otherwise specified.
All points of each failure envelope are normalized by dividing by the maximum absolute value.
The resulting dataset is randomly shuffled and split into training and test sets with an $8{:}2$ ratio.
For the adaptive sampling results reported in Section~\ref{sec:results:adaptive}, the same parameter bounds are used for the solver inputs $(P_1,P_2,P_3,P_4)$ during the Latin hypercube sampling (LHS) \cite{mckay1979comparison} procedure.
In the adaptive sequential sampling algorithm, normalization is performed using min--max scaling to $[-1,1]$ together with absolute-max scaling, with scaling factors computed only from the initial dataset.
As new samples are sequentially appended during the adaptive sampling procedure, they are normalized using this fixed scaling rule, and the scaler is not updated thereafter.
For dataset comparison, each dataset is independently normalized to $[-1,1]$ using min--max and absolute-max scaling during training.

\subsection{Finite Difference Approximation of Signed Curvature}
\label{sec:results:FD}

In this section, we compare the FD and AD schemes for evaluating the signed curvature in \cref{eq:curvature}.
We consider the practical setting in which the physics-informed loss term of the neural operator is evaluated in batch, and examine how the computational cost scales with batch size.
The comparison is made in terms of wall-clock time under controlled evaluation settings.
The following choices are adopted throughout the comparisons in this section.
First, the evaluation points are prescribed explicitly for each scheme so that all timing results are obtained under fully specified evaluation settings.
Second, both schemes are implemented in a vectorized manner so that the timing comparison reflects the computational structure of the two curvature-evaluation strategies rather than avoidable overhead from non-vectorized implementations.
For FD, the vectorized implementation follows directly from the finite difference stencils on the evaluation grid.
For AD, we employ the batched formulation described in \cref{sec:forward-oper-net}, which avoids tiling the inputs to the trunk network.
Third, all wall times reported in this section correspond to CPU time.

We first compare the forward evaluation time of the two schemes.
Both schemes are evaluated on the same equispaced grid in $[0,2\pi)$ for each configuration, across batch sizes $1$, $4$, $16$, $64$, $\dots$, and  $4{,}096$ 
and with dummy input parameters $\boldsymbol{\xi}$ that are chosen as zero vectors $\boldsymbol{0}$.
Each timing experiment is repeated $100$ times after $5$ warm-up iterations.

\begin{figure}[htbp]
    \centering
    \begin{subfigure}{0.4\textwidth}
        \centering
        \includegraphics[width=\textwidth]{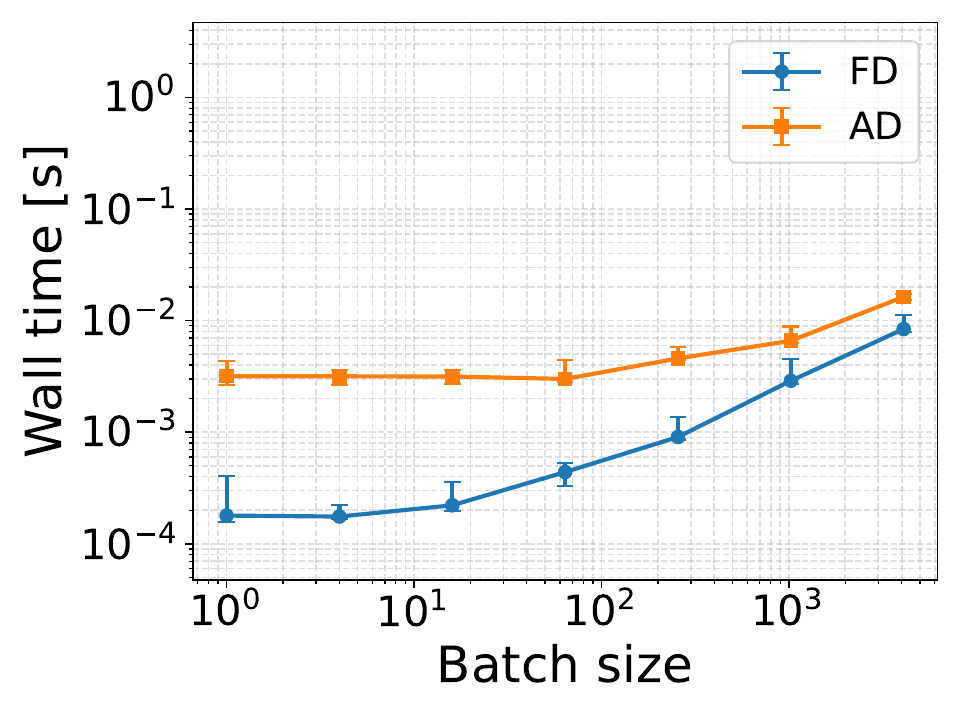}
        \caption{}\label{fig:bench_forward_a}
    \end{subfigure}
    \begin{subfigure}{0.4\textwidth}
        \centering
        \includegraphics[width=\textwidth]{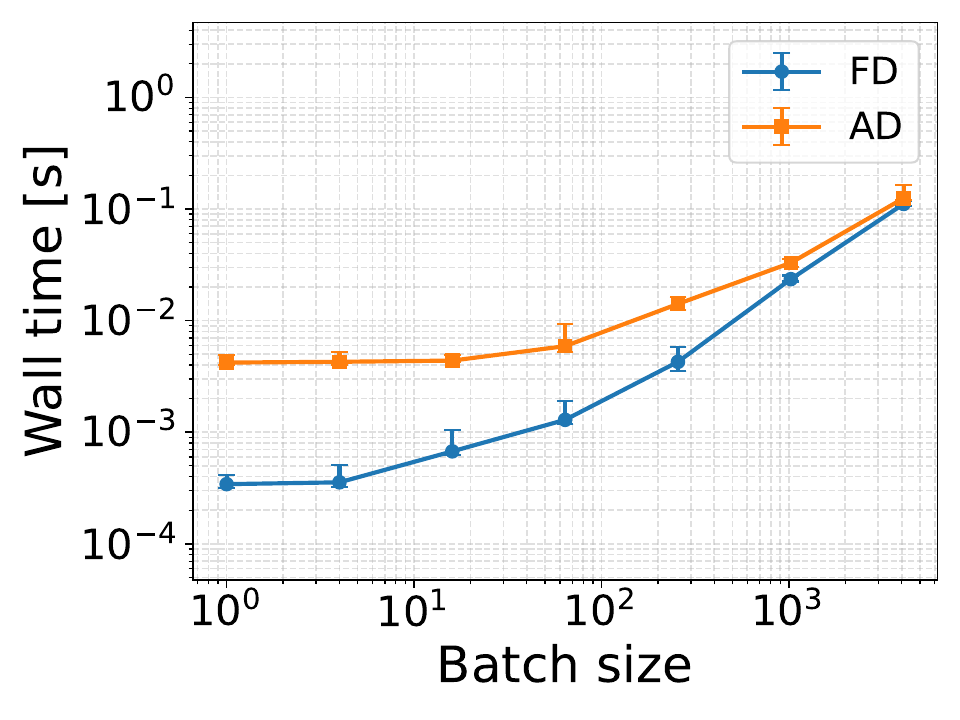}
        \caption{}\label{fig:bench_forward_b}
    \end{subfigure}
    \hfill
    \caption{
        Comparison of the computational time of the FD and AD schemes across different batch sizes.
        Results for
        (\subref{fig:bench_forward_a}) $100$ and
        (\subref{fig:bench_forward_b}) $1{,}000$ evaluation points are shown.
        Markers indicate the median over $100$ trials, where error bars denote the $25\%$ and $75\%$ quantiles.
    }
    \label{fig:bench_forward}
\end{figure}

The results are shown in \cref{fig:bench_forward}.
For the case of $100$ evaluation points, shown in \cref{fig:bench_forward_a}, AD is consistently slower than FD across all batch sizes considered.
Moreover, under the present implementation and workload, the measured AD runtime exhibits only weak dependence on batch size over the tested range, whereas the FD runtime shows a more noticeable increase.
For the case of $1{,}000$ evaluation points, shown in \cref{fig:bench_forward_b}, the wall times of both FD and AD increase relative to the case with $100$ evaluation points.
Similarly with the case of $100$ evaluation points, the increase with batch size remains less pronounced for AD than for FD.
As the batch size grows, the FD wall time approaches that of AD, and at batch size $4{,}096$ the two become comparable.

The similar tendency can be found in the work of \cite{jiang2023applications}, where FD was also found to be preferable to AD in terms of wall clock time for an MLP based formulation of the two-dimensional steady incompressible Navier--Stokes equations.
However, that study did not consider a neural operator setting.
Therefore, the comparison is most directly relevant to our case with batch size $1$, for which FD is likewise faster than AD for the grid sizes considered.
In the present DeepONet setting, by contrast, AD benefits from efficient vectorized differentiation across the additional data dimension beyond the coordinate support, which partly offsets the computational disadvantage often associated with AD.
Nevertheless, for the problem sizes most relevant in our setting, where the number of evaluation points is typically moderate, around $144$ rather than $1{,}000$, FD remains clearly preferable in terms of wall clock time.
% Overall, these observations indicate that, in the present benchmark setting, the AD runtime is less sensitive than the FD runtime to increases in batch size, 
% while the more favorable scheme in wall clock time depends on the specific problem setting.
Overall, these observations indicate that, in the present benchmark setting, the AD runtime is less sensitive than the FD runtime to increases in batch size, while for the shown cases FD tends to be faster in the inference of curvatures.

We next compare the training-time behavior of the FD and AD schemes.
While the forward-time comparison isolates the cost of curvature evaluation, it does not fully characterize the computational efficiency of the two schemes during training.
In particular, the two approaches may differ not only in per-iteration cost, but also in the resulting training dynamics, since FD evaluates the signed curvature through a finite difference approximation whereas AD evaluates it by exact differentiation.
Moreover, because the training objective in \cref{eq:loss_ftn} couples the data-misfit and curvature terms, the effect of the curvature-evaluation scheme must be assessed at the level of the full training process.

The comparison is performed for batch sizes $100$ and $1{,}000$, corresponding to full-batch training on reduced datasets of those sizes.
For each batch size, the training batch is selected randomly from the full dataset, and the curvature penalty term is evaluated on the fixed grid for both schemes.
For visualization and a consistent comparison of non-convexity, the fraction of negative curvature values is evaluated by querying both models on a common grid of $1{,}000$ equispaced points over the same interval and then computing the signed curvature using AD.
Accordingly, the fraction of negative curvature values is defined as the ratio of the number of points with negative signed curvature to the total number of queried points across the batch.

\begin{figure}[htbp]
    \centering
    \begin{subfigure}{0.4\textwidth}
        \centering
        \includegraphics[width=\textwidth]{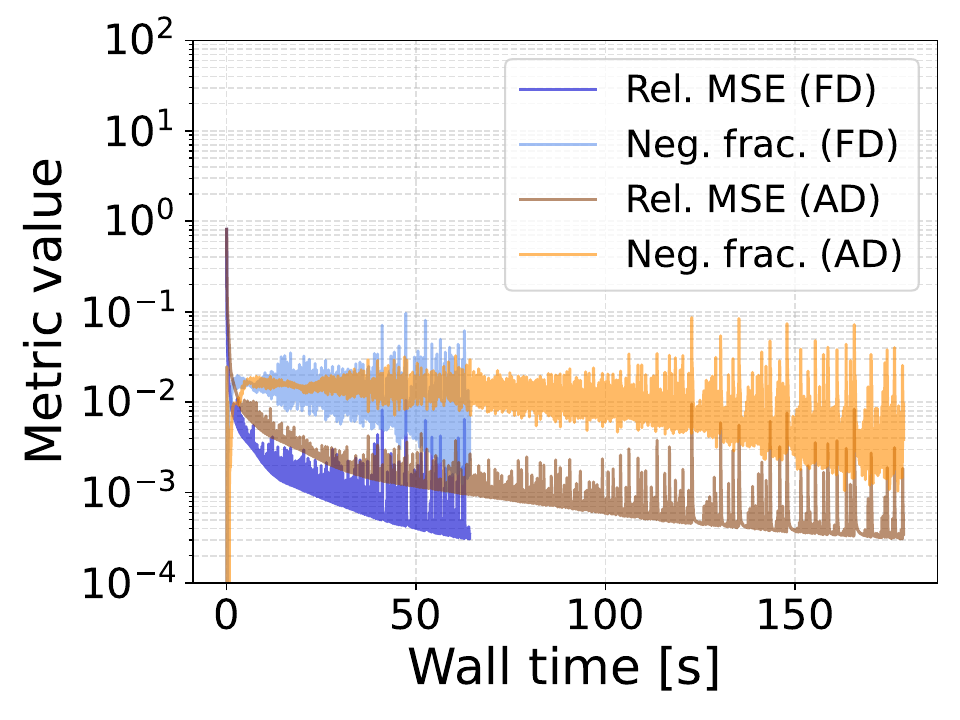}
        \caption{}\label{fig:bench_training_a}
    \end{subfigure}
    \begin{subfigure}{0.4\textwidth}
        \centering
        \includegraphics[width=\textwidth]{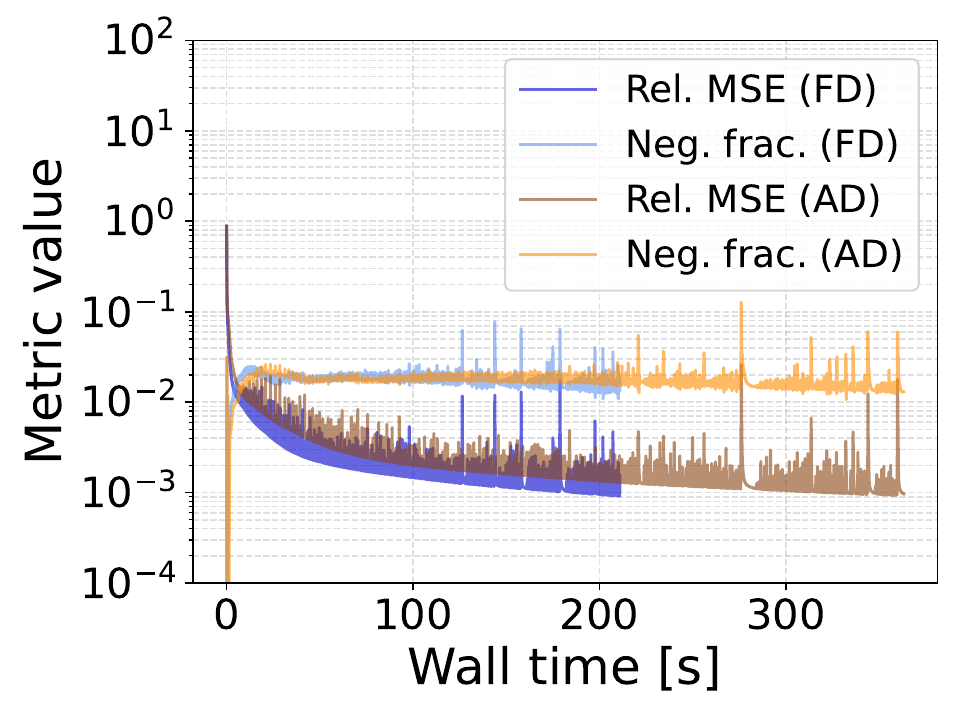}
        \caption{}\label{fig:bench_training_b}
    \end{subfigure}
    \hfill
    \caption{
        Histories of the relative MSE and the fraction of negative curvature values, comparing the FD and AD schemes for different batch sizes.
        (\subref{fig:bench_training_a}) Batch size $100$ and
        (\subref{fig:bench_training_b}) $1{,}000$.
    }
    \label{fig:bench_training}
\end{figure}

The resulting histories of the relative MSE and the fraction of negative curvature values, plotted against wall-clock time, are shown in \cref{fig:bench_training}.
As shown in \cref{fig:bench_training_a}, for batch size $100$, the two schemes exhibit broadly comparable values in both metrics over the course of training, without a substantial qualitative difference in the attained training state when viewed against the number of epochs.
However, FD reaches comparable levels of the relative MSE and the fraction of negative curvature values in significantly less wall-clock time than AD.
A similar trend is observed in \cref{fig:bench_training_b} for batch size $1{,}000$.
Again, after the same number of training epochs, the data-misfit term and the overall fraction of non-convexity remain broadly comparable between the two schemes.
One notable difference is that, whereas the wall-clock time of FD increases substantially---by considerably more than a factor of two---as the batch size grows from $100$ to $1{,}000$, the corresponding increase for AD is much less pronounced, remaining close to a factor of two.
This suggests that, in the present AD formulation, the computational cost is more dominated by components that scale weakly with the batch size (e.g., evaluation of the curvature penalty term), whereas the FD cost scales more directly with the batch size.
These observations are consistent with those from the curvature evaluation setting in \cref{fig:bench_forward}.
Nevertheless, FD remains faster than AD in terms of wall-clock time for the settings shown.
Overall, these results suggest that FD provides a practical alternative to AD for evaluating the curvature-based regularization during training, yielding comparable training behavior while reducing computational cost.
The settings adopted in Section~\ref{sec:results:surrogate} lie between the two benchmark scenarios considered here, for which FD remains favorable.

\subsection{Physics-informed Neural Operator}\label{sec:results:surrogate}
Our objective in this section is to demonstrate that the forward neural operator can accurately learn the underlying operator from irregularly sampled data, while the physics-informed loss enforcing convexity promotes mechanically admissible responses. As a result, the surrogate model serves \textit{not only as a ``predictor'', but also as a ``corrective'' model for the micromechanical simulator}.

To emulate incomplete and irregular observations, each sample is randomly masked using a ratio drawn from $\mathcal{U}(0.2,0.5)$, yielding point cloud representations with varying resolution (different number of sampled points on each failure envelope).
Here, two surrogate models are trained under different objective settings.
One model is trained by minimizing a composite objective consisting of both a data-misfit term \cref{eq:loss_rel_mse} and a curvature-based regularization \cref{eq:curvature_penalty},
while the other is trained by minimizing an objective consisting solely of the data-misfit term.
All experiments in this section and Section~\ref{sec:results:inverse} use these irregularly sampled point cloud data. 
Additional details of the training are provided in Appendix~\ref{apdx:hyperparams}.

\begin{figure}[htbp]
    \centering
    \begin{subfigure}{0.4\textwidth}
        \centering
        \includegraphics[width=\textwidth]{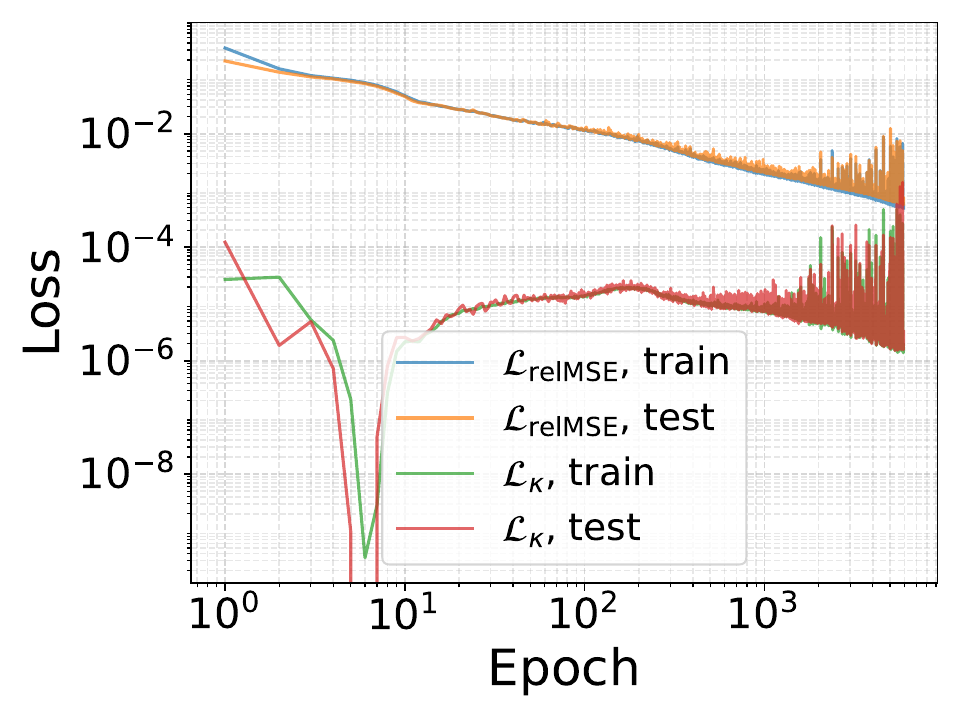}
        \caption{}\label{fig:pr1_loss_a}
    \end{subfigure}
    \begin{subfigure}{0.4\textwidth}
        \centering
        \includegraphics[width=\textwidth]{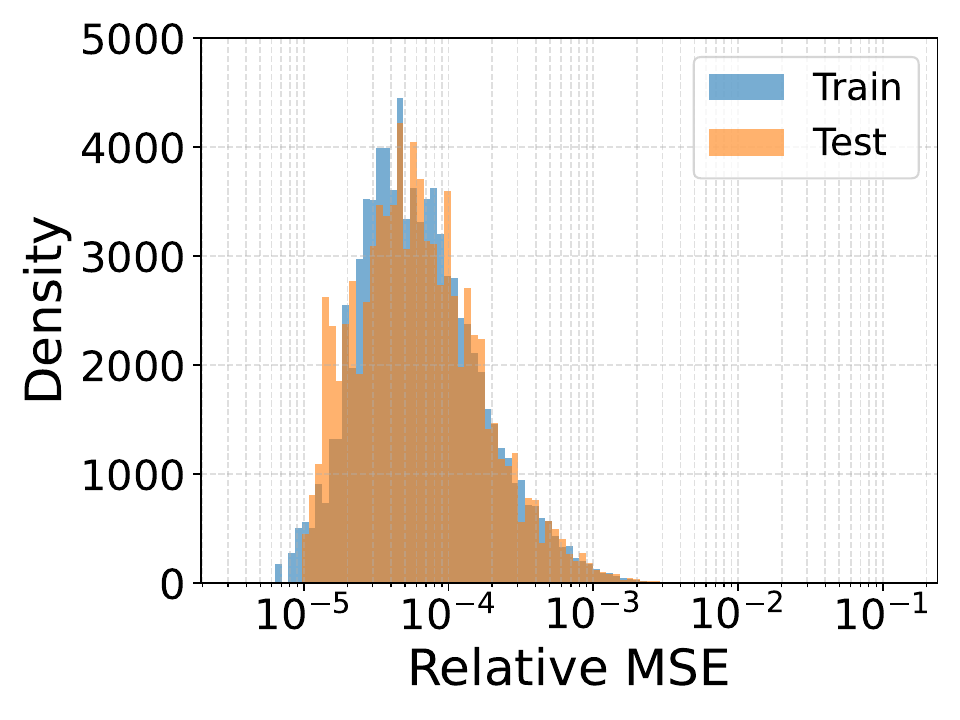}
        \caption{}\label{fig:pr1_loss_b}
    \end{subfigure}
    \begin{subfigure}{0.4\textwidth}
        \centering
        \includegraphics[width=\textwidth]{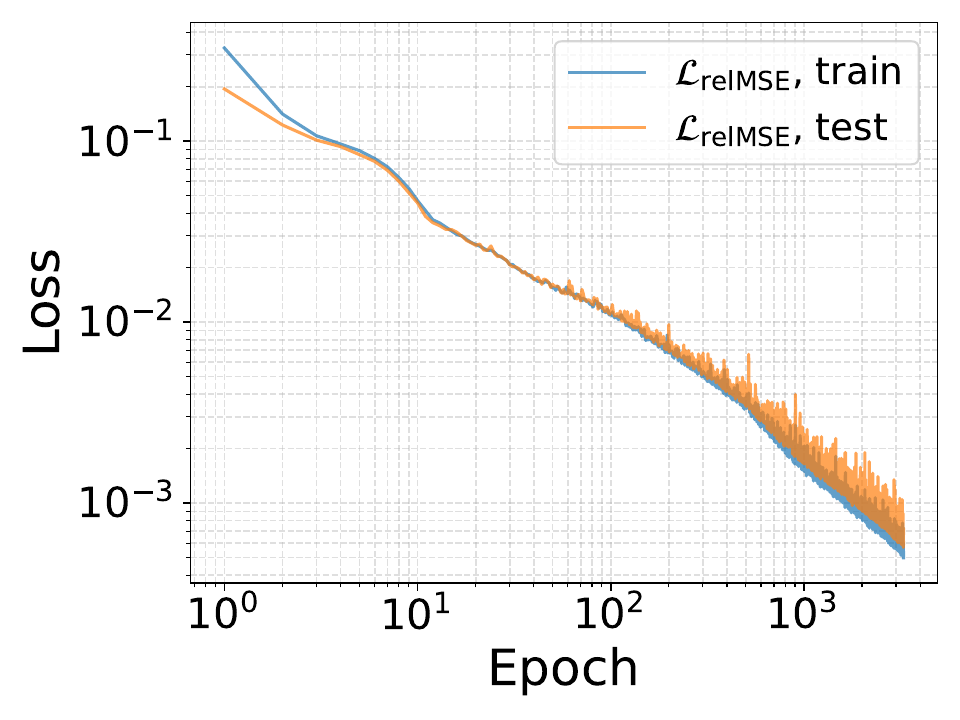}
        \caption{}\label{fig:pr1_loss_c}
    \end{subfigure}
    \begin{subfigure}{0.4\textwidth}
        \centering
        \includegraphics[width=\textwidth]{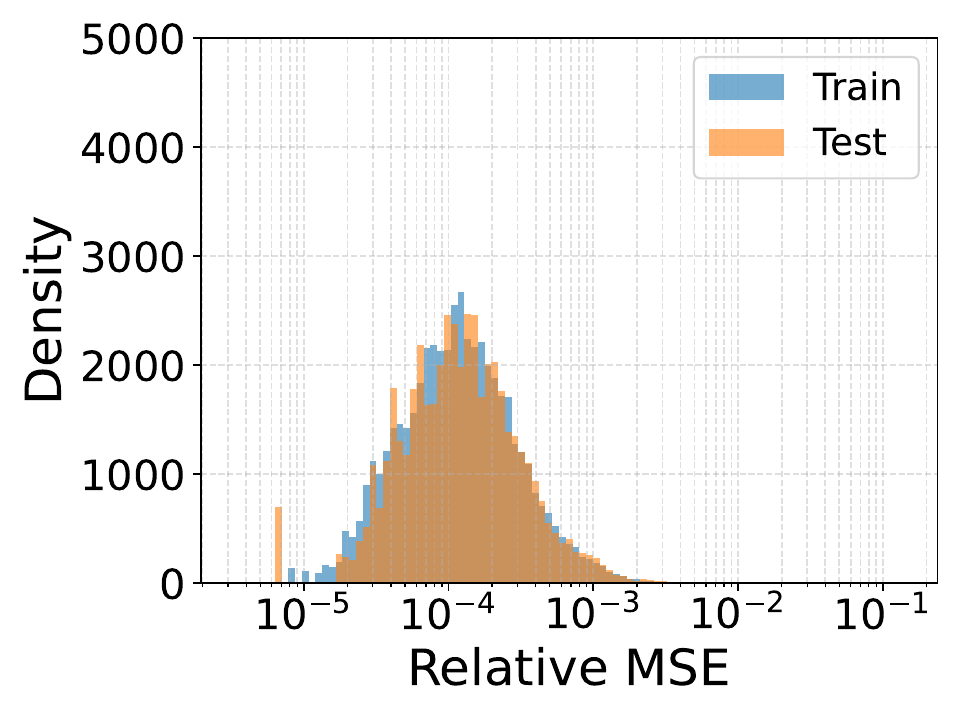}
        \caption{}\label{fig:pr1_loss_d}
    \end{subfigure}
    \hfill
    \caption{
    Training curves and error distributions of the surrogate models.
    (\subref{fig:pr1_loss_a}) Training and test loss histories with the curvature penalty term,
    (\subref{fig:pr1_loss_b}) histograms of relative MSE on train and test data with the curvature penalty term,
    (\subref{fig:pr1_loss_c}) training and test loss histories without the curvature penalty term,
    (\subref{fig:pr1_loss_d}) and histograms of relative MSE on train and test data without the curvature penalty term.
    }
    \label{fig:pr1_loss}
\end{figure}

\cref{fig:pr1_loss} summarizes the training behavior of the surrogate models.
As shown in \cref{fig:pr1_loss_a,fig:pr1_loss_b}, the training and test losses remain comparable throughout training, indicating limited overfitting under the present regularization.
This observation is further supported by \cref{fig:pr1_loss_c,fig:pr1_loss_d}, where the distributions of the training and test errors are similar in each case.
In addition, \cref{fig:pr1_loss_a} shows that the curvature penalty term $\mathcal{L}_{\kappa}$ is initially zero; as the model fits the data, negative curvature regions emerge and are penalized, while the overall loss exhibits a decreasing trend.
This behavior suggests that the prior motivated by Drucker’s postulate effectively biases the surrogate outputs toward convex shapes.

\begin{figure}[htbp]
    \centering
    \begin{subfigure}{0.35\textwidth}
        \centering
        \includegraphics[width=\textwidth]{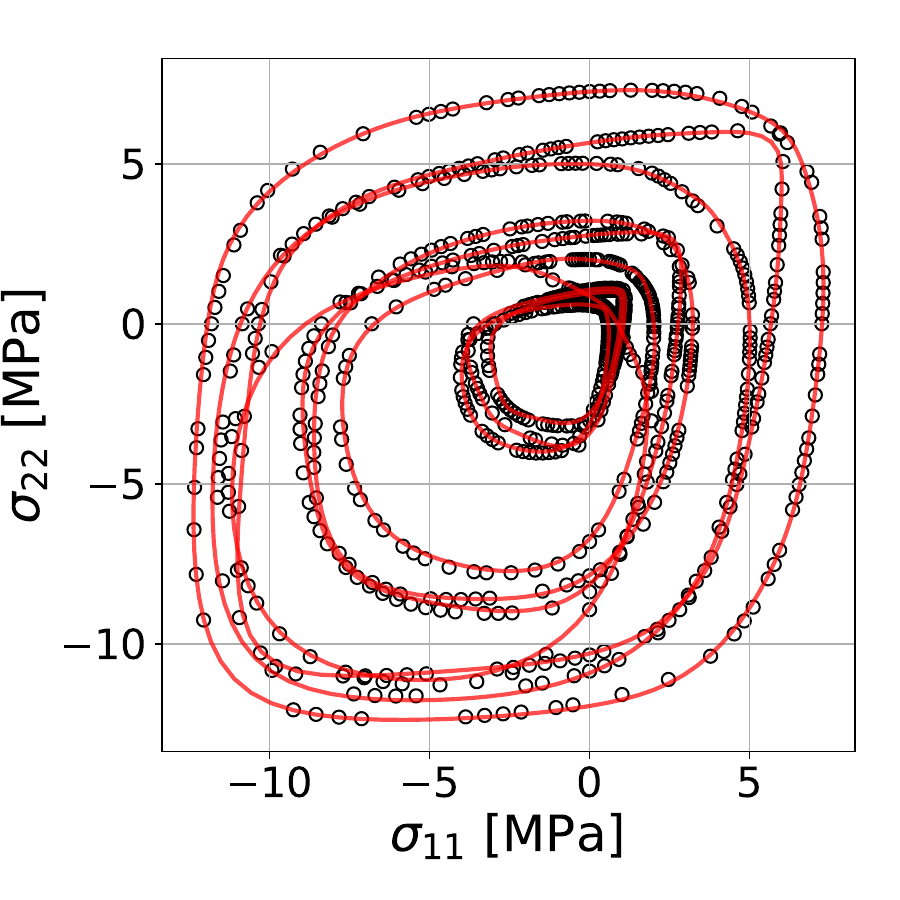}
        \caption{}\label{fig:pr1_surrogate_a}
    \end{subfigure}
    \begin{subfigure}{0.35\textwidth}
        \centering
        \includegraphics[width=\textwidth]{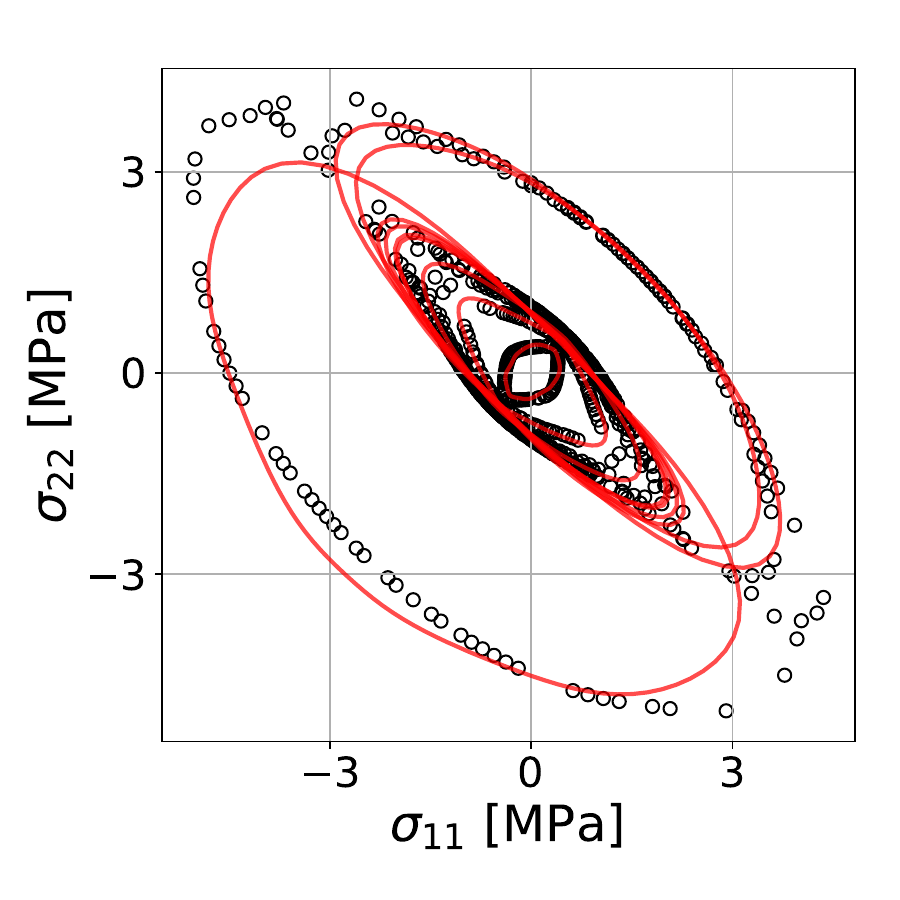}
        \caption{}\label{fig:pr1_surrogate_b}
    \end{subfigure}
    \begin{subfigure}{0.35\textwidth}
        \centering
        \includegraphics[width=\textwidth]{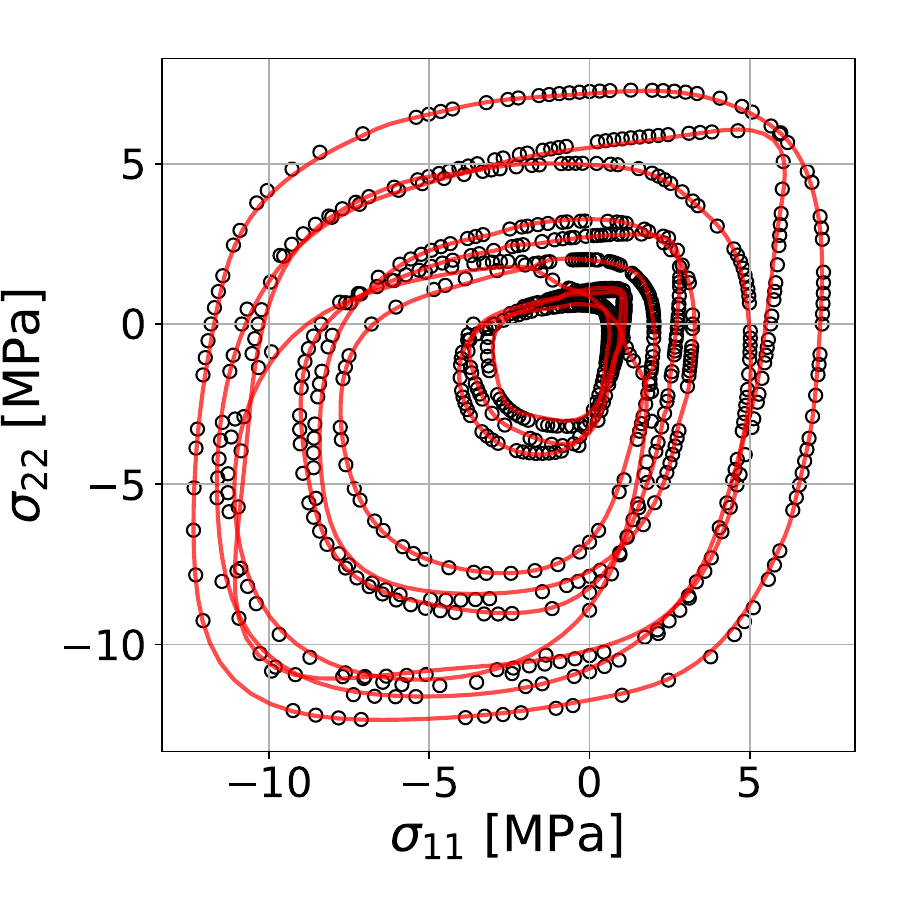}
        \caption{}\label{fig:pr1_surrogate_c}
    \end{subfigure}
    \begin{subfigure}{0.35\textwidth}
        \centering
        \includegraphics[width=\textwidth]{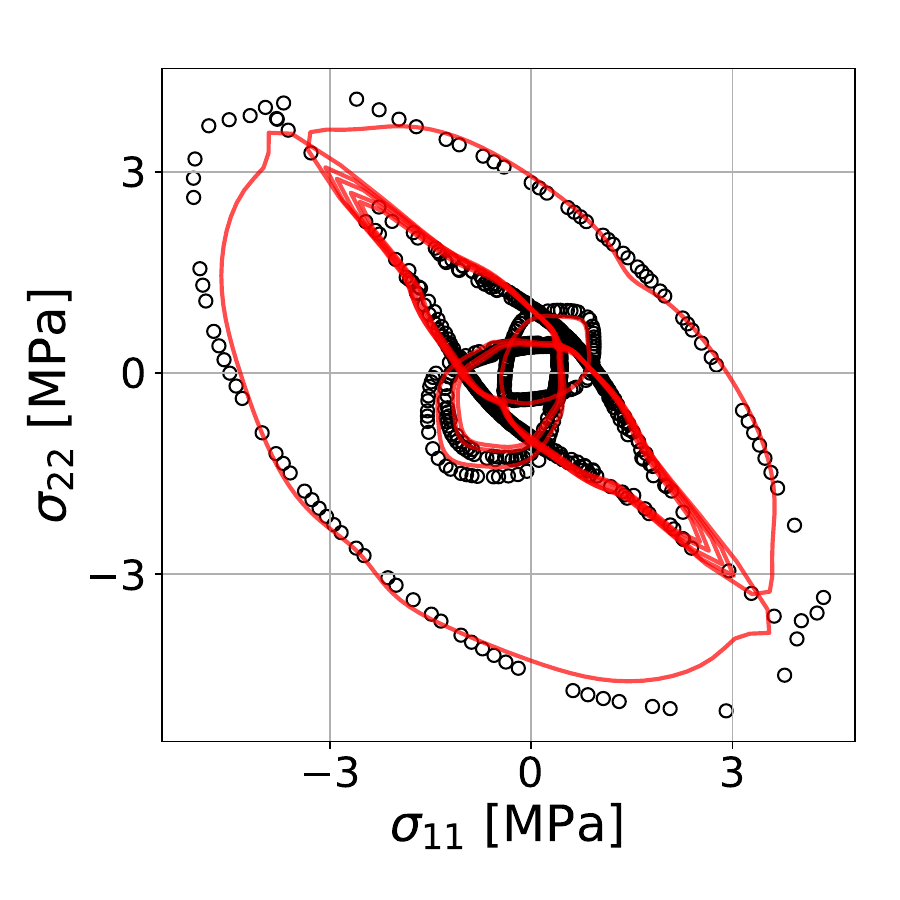}
        \caption{}\label{fig:pr1_surrogate_d}
    \end{subfigure}
    \hfill
    \caption{
    Visualization of surrogate model predictions on the test data.
    Predictions of the model trained with the curvature penalty are shown for (\subref{fig:pr1_surrogate_a}) ten randomly selected cases and
    (\subref{fig:pr1_surrogate_b}) ten cases with the largest relative MSE.
    Predictions of the model trained without the curvature penalty are shown for (\subref{fig:pr1_surrogate_c}) the same ten cases as in (\subref{fig:pr1_surrogate_b}) and
    (\subref{fig:pr1_surrogate_d}) ten cases with the largest relative MSE.
    }
    \label{fig:pr1_surrogate}
\end{figure}

\begin{figure}[htbp]
    \centering
    \begin{subfigure}{0.3\textwidth}
        \centering
        \includegraphics[width=\textwidth]{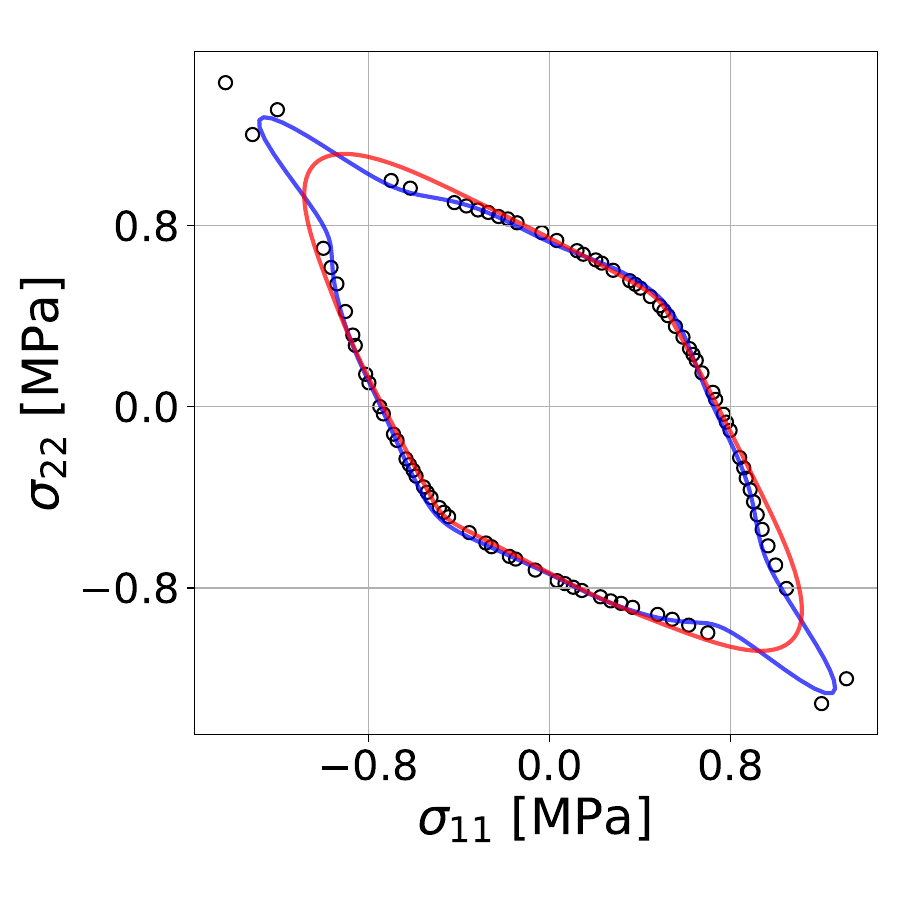}
    \end{subfigure}
    \begin{subfigure}{0.3\textwidth}
        \centering
        \includegraphics[width=\textwidth]{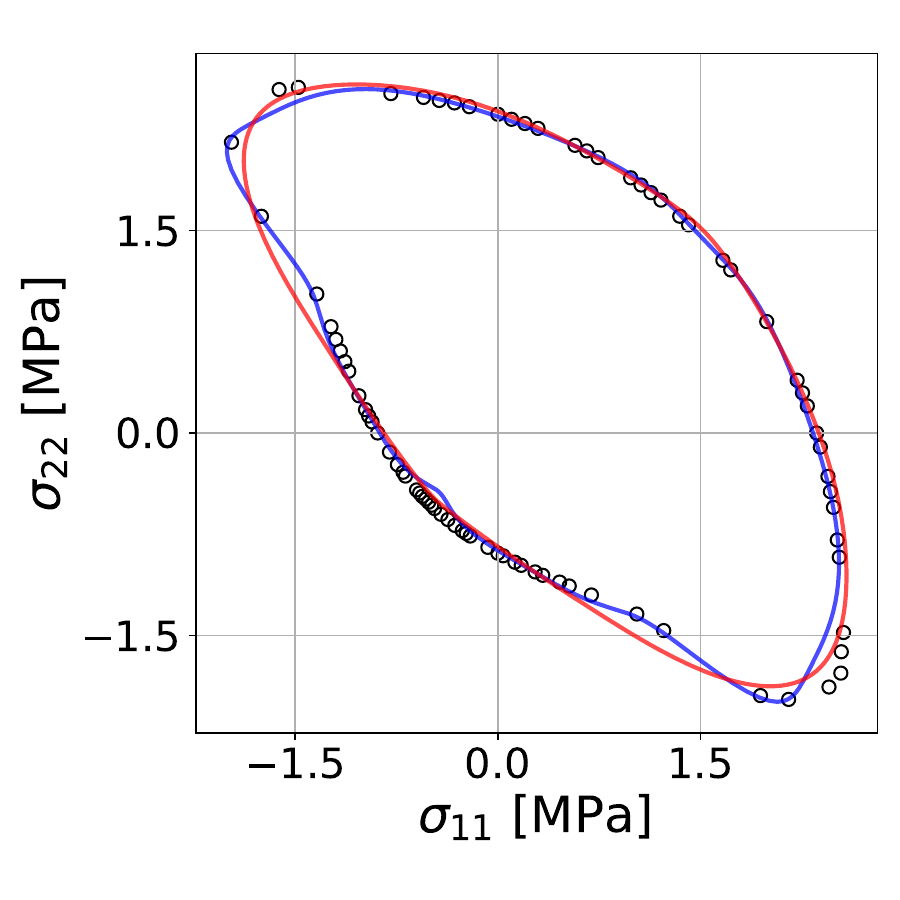}
    \end{subfigure}
    \hfill
    \caption{
    Representative test cases illustrating the predicted failure envelopes. Blue curves denote the surrogate model trained without curvature regularization, red curves denote the model with the proposed curvature regularization, and black open circles denote the reference data.
    }
    \label{fig:pr1_nonconvex}
\end{figure}

The predicted failure envelopes from the surrogate model are depicted in Figure~\ref{fig:pr1_surrogate}.
Overall, the surrogate predictions closely track the reference envelopes.
As shown in \cref{fig:pr1_surrogate_a,fig:pr1_surrogate_b}, the model trained with the curvature penalty exhibits consistently convex behavior across varying scales.
\cref{fig:pr1_surrogate_c} shows that the model trained without the penalty also produces largely convex outputs in many regions.
However, \cref{fig:pr1_surrogate_d} indicates that, without the curvature penalty, the model often produces locally non-convex outputs.
These non-convexities are more clearly observed in \cref{fig:pr1_nonconvex}, where the surrogate model trained with the curvature penalty term remains nearly convex overall for the representative cases, whereas the model trained without the penalty exhibits a noticeable fraction of locally non-convex regions.

\begin{figure}[htbp]
    \centering
    \begin{subfigure}{0.3\textwidth}
        \centering
        \includegraphics[width=\textwidth]{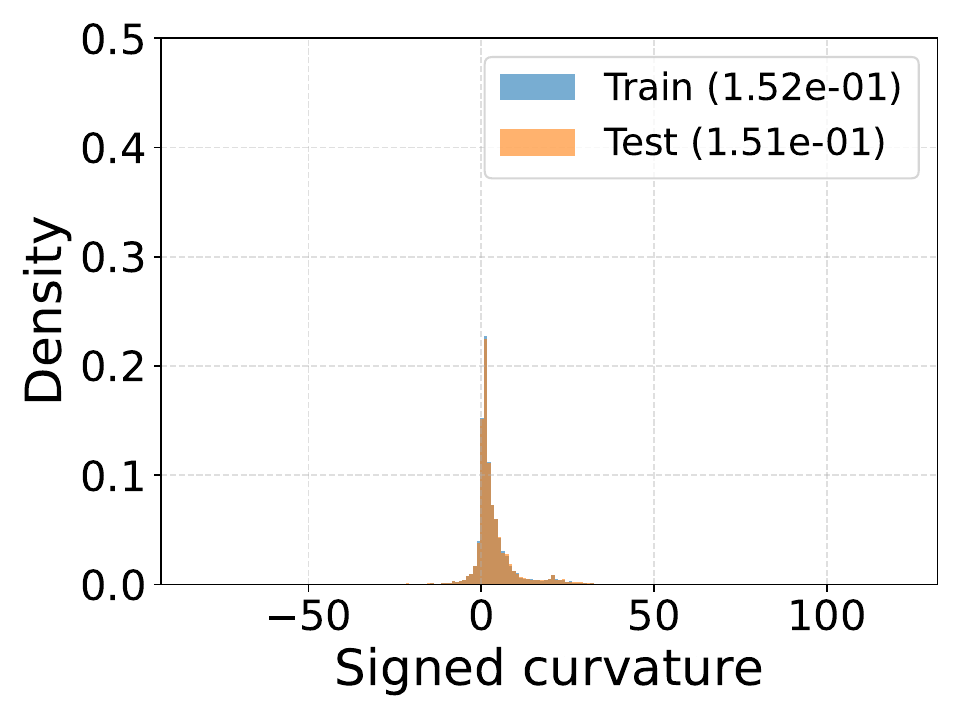}
        \caption{}\label{fig:pr1_hist_curvature_a}
    \end{subfigure}
    \begin{subfigure}{0.3\textwidth}
        \centering
        \includegraphics[width=\textwidth]{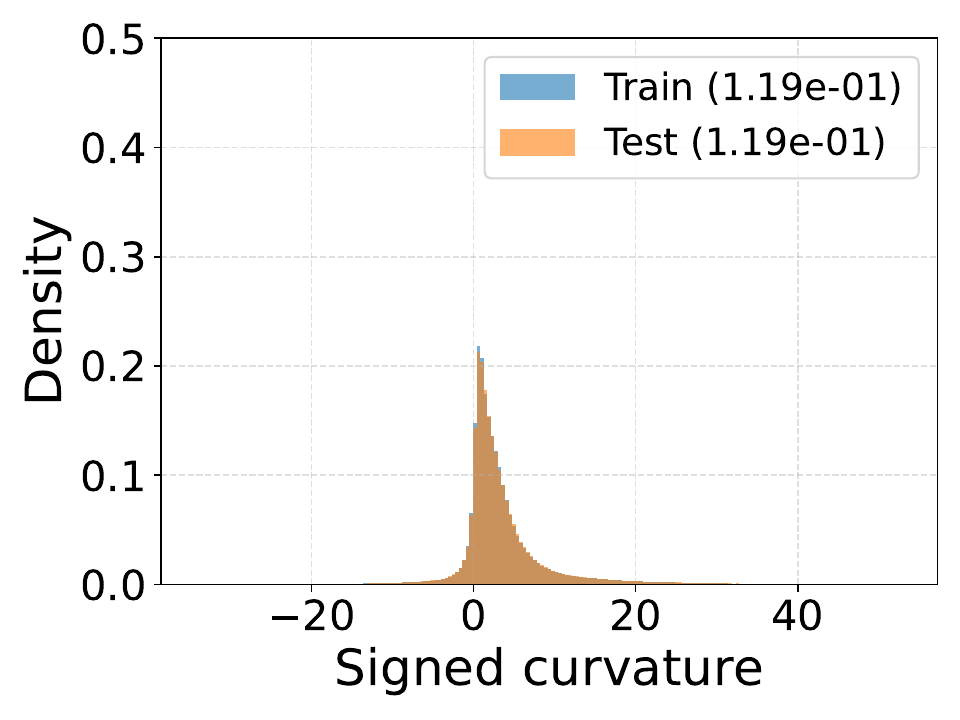}
        \caption{}\label{fig:pr1_hist_curvature_b}
    \end{subfigure}
    \begin{subfigure}{0.3\textwidth}
        \centering
        \includegraphics[width=\textwidth]{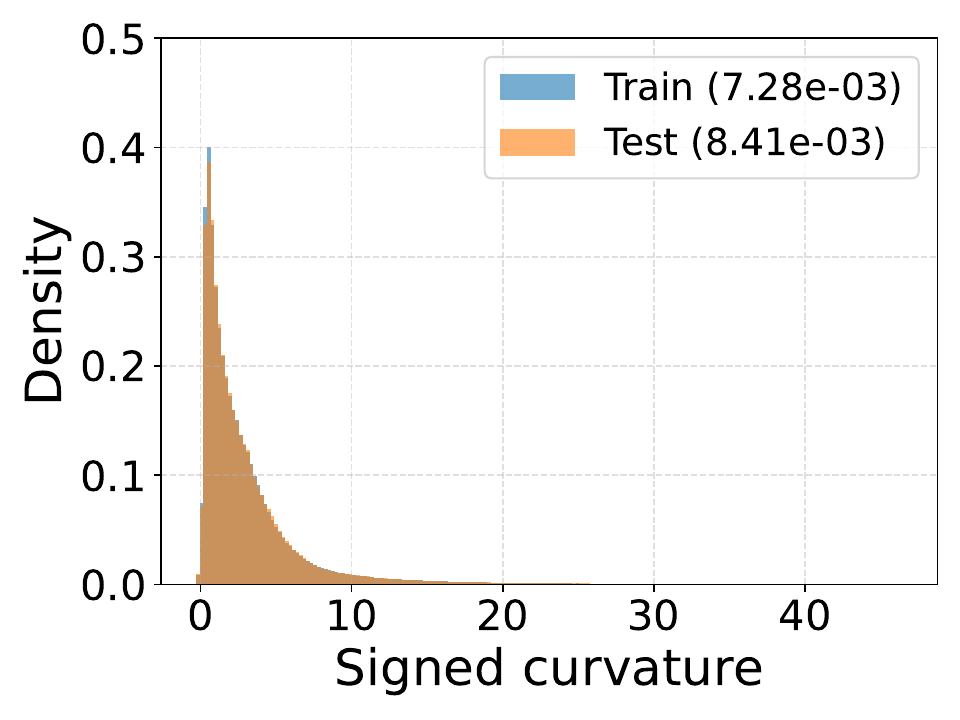}
        \caption{}\label{fig:pr1_hist_curvature_c}
    \end{subfigure}
    \hfill
    \caption{
    (\subref{fig:pr1_hist_curvature_a}) Distribution of the signed curvature in the given data.
    Distributions of the signed curvature for the trained surrogate models,
    (\subref{fig:pr1_hist_curvature_b}) without the curvature penalty and
    (\subref{fig:pr1_hist_curvature_c}) with the curvature penalty.
    The values shown in parentheses indicate the fraction of negative signed curvature values.
    }
    \label{fig:pr1_hist_curvature}
\end{figure}

To quantify this behavior, we evaluate the outputs of the two surrogate models using the signed curvature defined in \cref{eq:curvature}.
For each latent parameter pair in both the training and test sets, the model is queried at $1{,}000$ equispaced values of $t \in [0,2\pi)$, and the signed curvature is then calculated using the formulation in \cref{eq:curvature} where the derivatives are evaluated by AD.
For the data, the signed curvature is evaluated using a central difference scheme.
For visualization, only the values between the $0.5\%$ and $99.5\%$ quantiles are shown.
The resulting distributions are presented in \cref{fig:pr1_hist_curvature}.
According to \cref{fig:pr1_hist_curvature_a}, approximately $15\%$ of the points in the given data exhibit local non-convexity.
In contrast, \cref{fig:pr1_hist_curvature_b} shows that the surrogate trained without the penalty exhibits a non-negligible amount of non-convex behavior, with approximately $12\%$ of the points having negative signed curvature.
This suggests that, even without explicit regularization, the surrogate partially smooths the non-convexity present in the data by structural bias of the DeepONet.
As shown in \cref{fig:pr1_hist_curvature_c}, across both the training and test cases, the surrogate trained with the curvature penalty remains nearly convex, with the fraction of locally non-convex points below approximately $1\%$.
These results highlight the importance of the curvature-based regularization in promoting convexified failure envelope shapes beyond merely enforcing proximity between the predicted and target point clouds.

\subsection{Inverse Identification}\label{sec:results:inverse}

In this section, we demonstrate the capability of the trained surrogate model to act as a differentiable solver for identifying microstructure configurations corresponding to a target failure envelope. During the inverse optimization, the surrogate mismatch is used as the objective to guide the search. However, it is not taken as the final measure of success. Once an optimal microstructure configuration is identified, it is evaluated using the high-fidelity micromechanical simulator, and the discrepancy between the resulting failure envelope and the target is used as the true performance metric.

We consider a target failure envelope from test cases not seen during training. The inverse optimization requires an initial guess and may admit multiple feasible solutions, with sensitivity to initialization. Fully random initialization may place the parameters in regions insufficiently explored by the forward model, where gradient information can be unreliable and lead to poor convergence.
To mitigate this, we randomly select ten initial parameter vectors from the training data and use them as initial guesses. Starting from these initializations, the parameters are optimized against the target envelope (see Appendix~\ref{apdx:hyperparams} for details).

\begin{figure}[htbp]
    \centering
    \begin{subfigure}{0.4\textwidth}
        \centering
        \includegraphics[width=\textwidth]{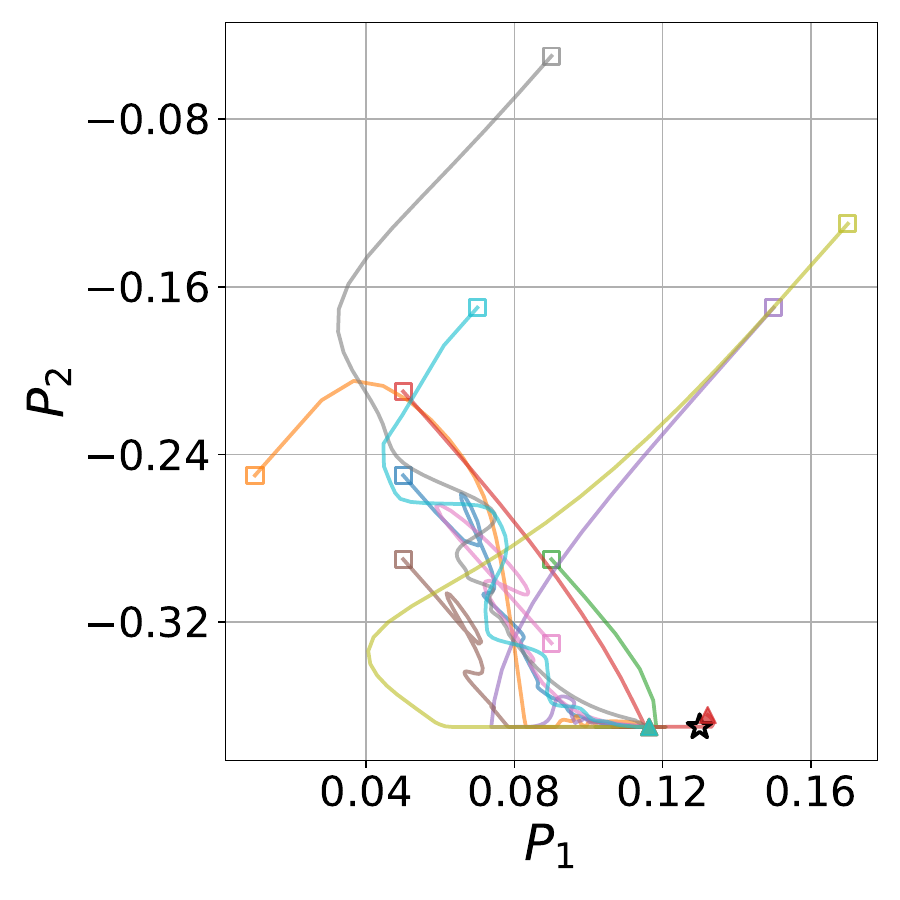}
        \caption{}\label{fig:pr1_inverse_a}
    \end{subfigure}
    \begin{subfigure}{0.4\textwidth}
        \centering
        \includegraphics[width=\textwidth]{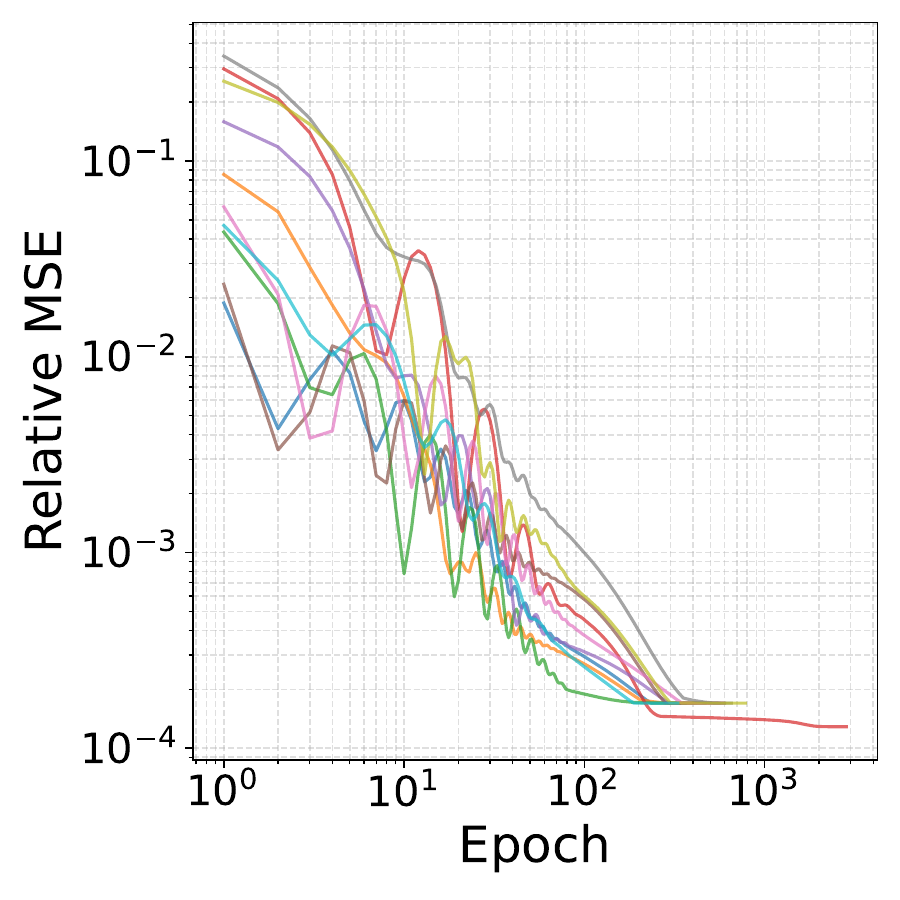}
        \caption{}\label{fig:pr1_inverse_b}
    \end{subfigure}
    \begin{subfigure}{0.4\textwidth}
        \centering
        \includegraphics[width=\textwidth]{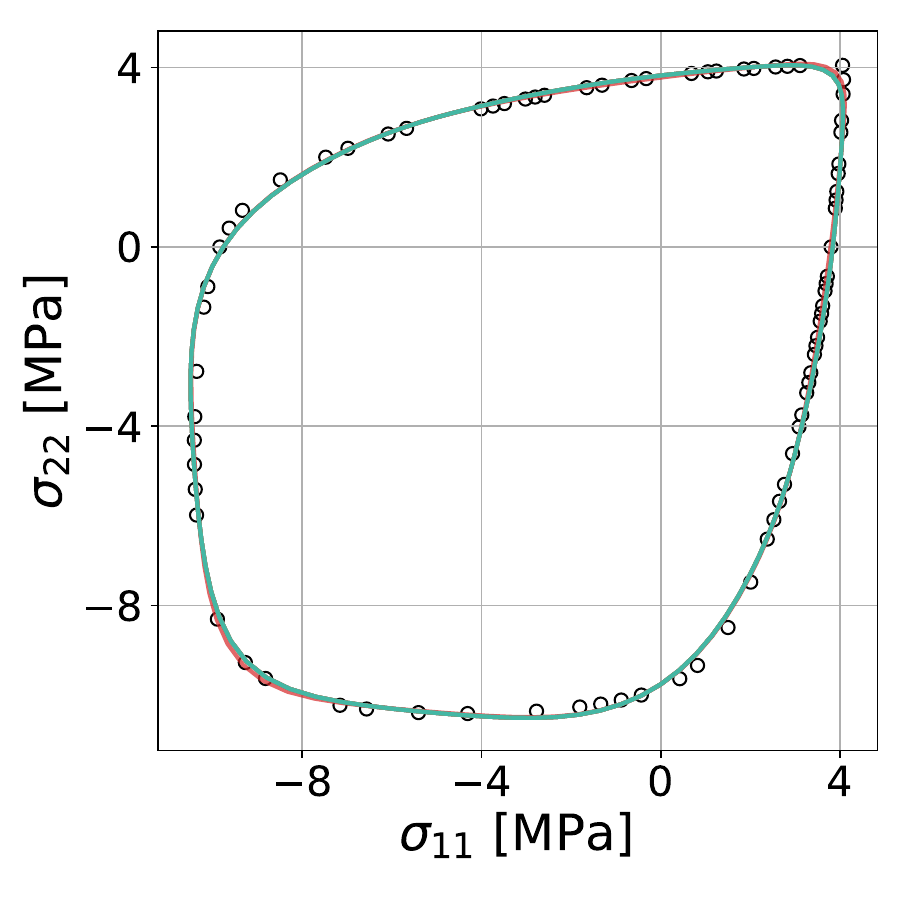}
        \caption{}\label{fig:pr1_inverse_c}
    \end{subfigure}
    \begin{subfigure}{0.4\textwidth}
        \centering
        \includegraphics[width=\textwidth]{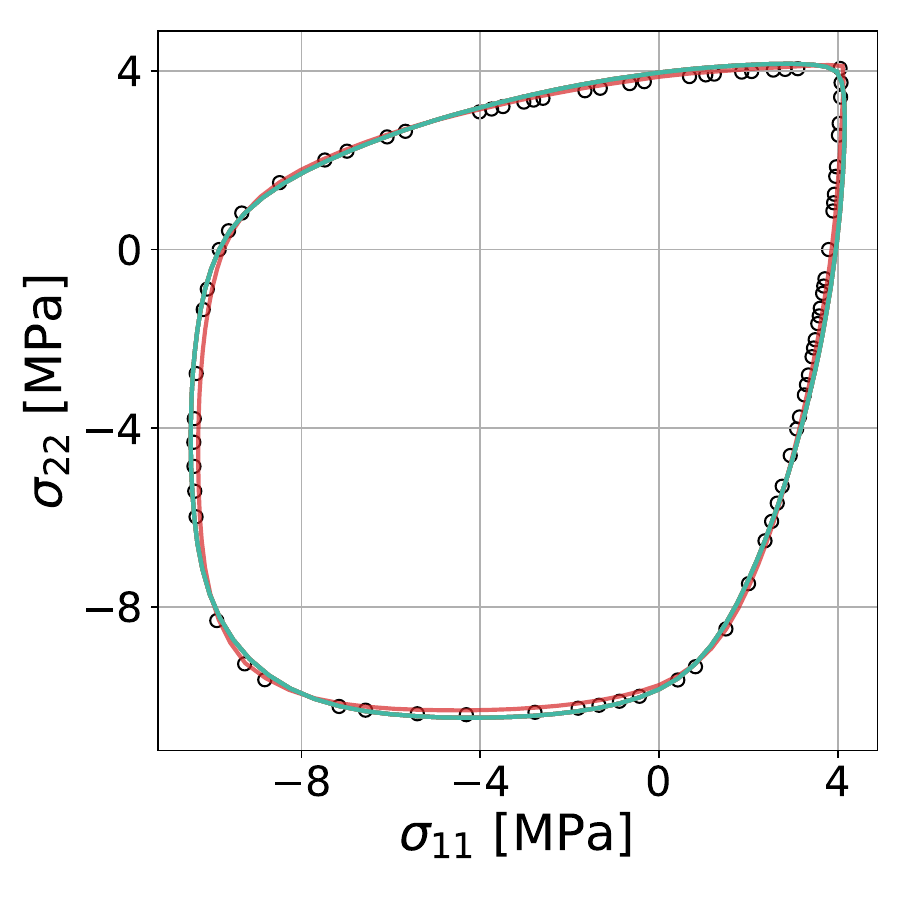}
        \caption{}\label{fig:pr1_inverse_d}
    \end{subfigure}
    \hfill
    \caption{
    Inverse identification of micromechanical parameters.
    (\subref{fig:pr1_inverse_a}) Ten trajectories of micromechanical parameters $P_1$ \& $P_2$ during the gradient descent. 
    Initial guesses are marked as open squares, converged parameters as triangles, and the target failure envelope as a star.
    (\subref{fig:pr1_inverse_b}) Loss histories of the ten trials.
    (\subref{fig:pr1_inverse_c}) Ten corresponding surrogate outputs as solid lines, and the data as black open circles.
    (\subref{fig:pr1_inverse_d}) Ten corresponding micromechanical solver outputs as solid lines, and the data as black open circles.
    }
    \label{fig:pr1_inverse}
\end{figure}

The inverse-identification results are shown in Fig.~\ref{fig:pr1_inverse}. In Fig.~\ref{fig:pr1_inverse_a}, we show the optimization trajectories projected onto the $(P_1,P_2)$ plane for visualization purposes, while the full parameter space is four-dimensional. The trajectories converge to neighborhoods of the ground-truth parameter configuration, but not to a single identical solution.
Such ill-posed behavior is also observed in \cref{fig:pr1_inverse_b}, where individual trials do not necessarily converge to the lowest-loss solution and settle in local minima.
Nevertheless, the corresponding surrogate-predicted failure envelopes in Fig.~\ref{fig:pr1_inverse_c} remain very similar to one another and to the target data, despite the visible separation among the recovered parameter values.
To confirm that this behavior is not an artifact of the surrogate, the converged parameter sets are further evaluated using the original micromechanical solver.
As shown in Fig.~\ref{fig:pr1_inverse_d}, the solver outputs are also nearly indistinguishable, indicating that distinct micromechanical parameter configurations can produce closely similar failure envelopes.
These results further highlight the practical value of the proposed differentiable surrogate, which enables efficient discovery of such competing parameter configurations without repeatedly using the full micromechanical solver.

\begin{figure}[htbp]
    \centering
    \begin{subfigure}{0.4\textwidth}
        \centering
        \includegraphics[width=\textwidth]{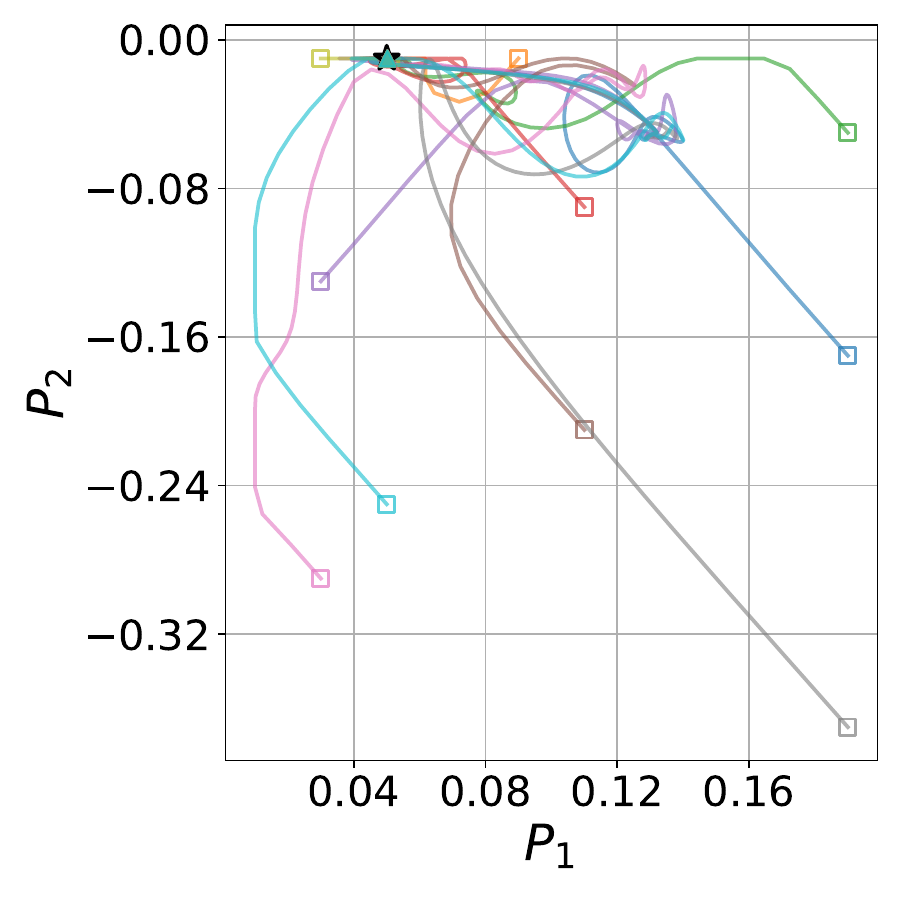}
        \caption{}\label{fig:pr1_inverse_noncovex_a}
    \end{subfigure}
    \begin{subfigure}{0.4\textwidth}
        \centering
        \includegraphics[width=\textwidth]{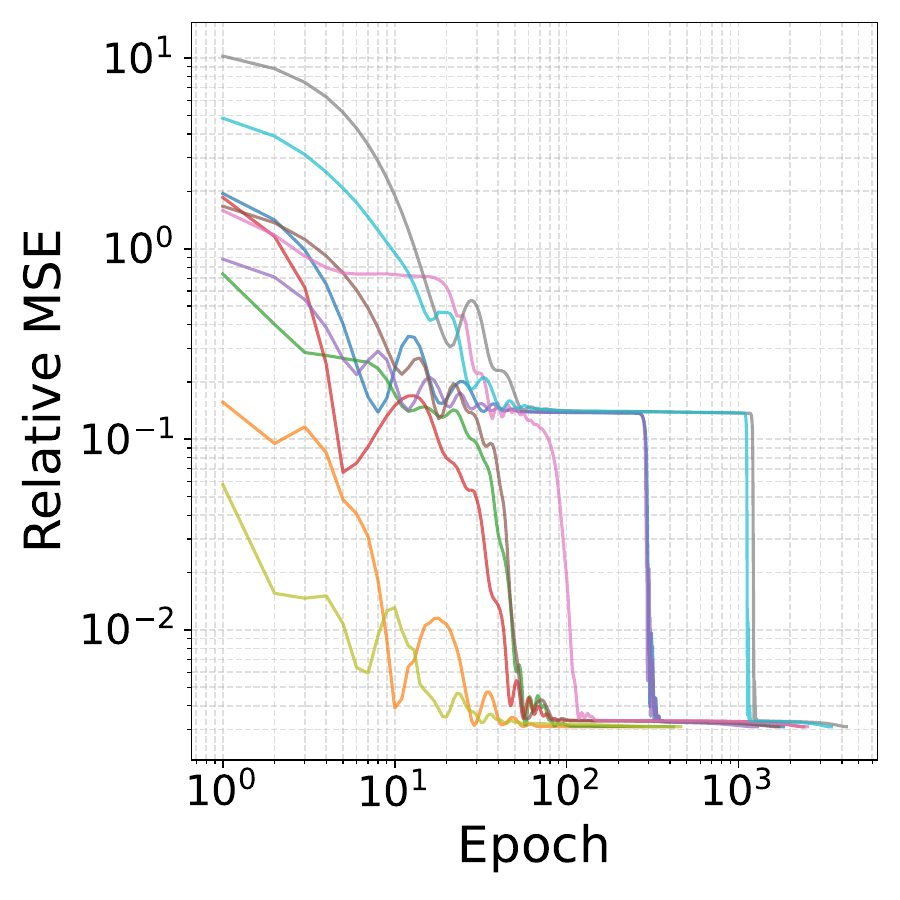}
        \caption{}\label{fig:pr1_inverse_noncovex_b}
    \end{subfigure}
    \begin{subfigure}{0.4\textwidth}
        \centering
        \includegraphics[width=\textwidth]{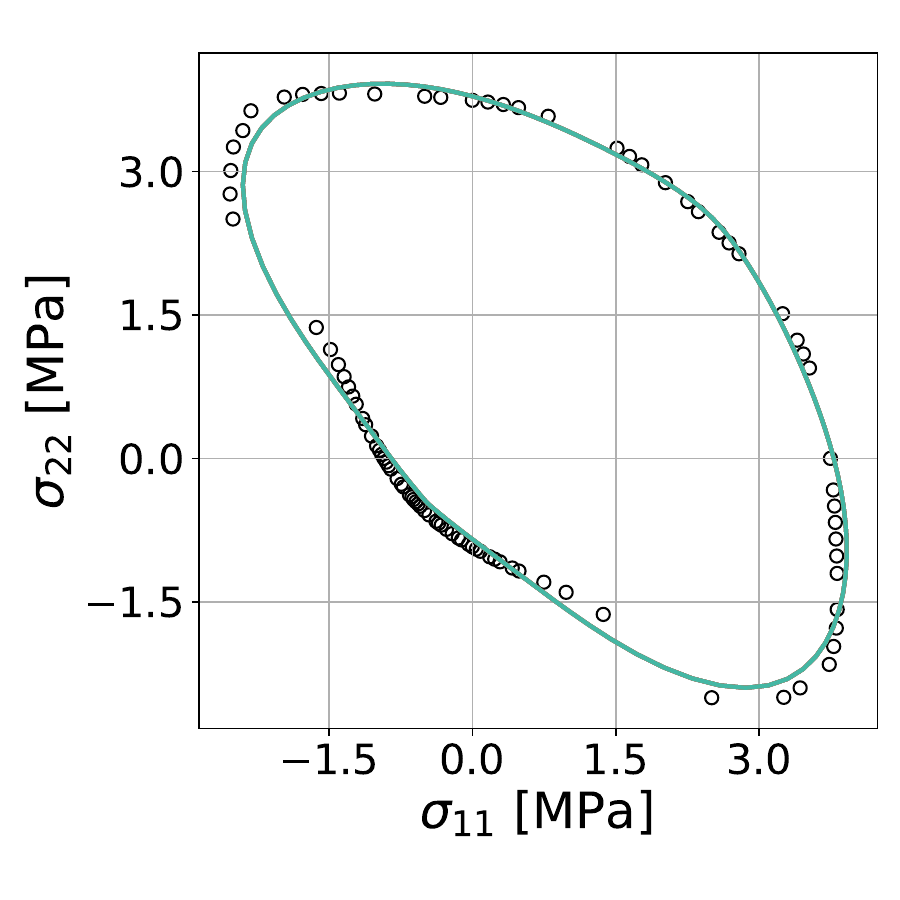}
        \caption{}\label{fig:pr1_inverse_noncovex_c}
    \end{subfigure}
    \begin{subfigure}{0.4\textwidth}
        \centering
        \includegraphics[width=\textwidth]{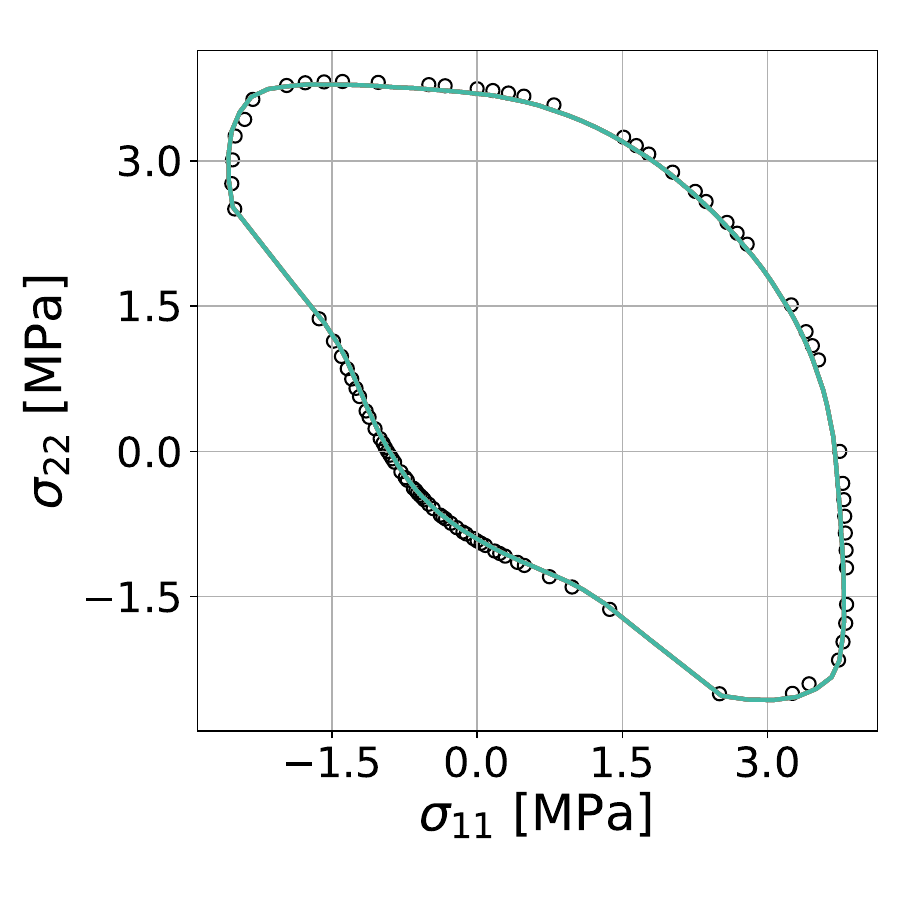}
        \caption{}\label{fig:pr1_inverse_noncovex_d}
    \end{subfigure}
    \hfill
    \caption{
    Inverse identification of micromechanical parameters on a representative case with non-convexity.
    (\subref{fig:pr1_inverse_noncovex_a}) Ten trajectories of micromechanical parameters $P_1$ \& $P_2$ during the gradient descent. 
    Initial guesses are marked as open squares, converged parameters as triangles, and the target failure envelope as a star.
    (\subref{fig:pr1_inverse_noncovex_b}) Loss histories of the ten trials.
    (\subref{fig:pr1_inverse_noncovex_c}) Ten corresponding surrogate outputs as solid lines, and the data as black open circles.
    (\subref{fig:pr1_inverse_noncovex_d}) Ten corresponding micromechanical solver outputs as solid lines, and the data as black open circles.
    }
    \label{fig:pr1_inverse_noncovex}
\end{figure}

We next consider another inverse identification setting, selected to examine the effect of the convexified surrogate output.
Specifically, we choose a representative test case for which the reference data exhibit local non-convex behavior.
The results are shown in \cref{fig:pr1_inverse_noncovex}.
As shown in \cref{fig:pr1_inverse_noncovex_a}, the ten trials converge closely to the true configuration in the $(P_1,P_2)$ projection.
The corresponding loss histories are reported in \cref{fig:pr1_inverse_noncovex_b}.
Trials initialized near the ground truth converge rapidly, while often others exhibit a prolonged plateau before successfully descending to the final solution.
The predicted failure envelopes corresponding to the converged parameters are shown in \cref{fig:pr1_inverse_noncovex_c}.
Since the identified parameters lie close to one another, the surrogate outputs are nearly indistinguishable. Importantly, while the inferred parameters are close to the true values, the surrogate output corresponds to a convex approximation of the target curve, providing further empirical evidence that the surrogate learns the closest convexification of the target. To further assess this behavior, the inferred parameter sets are evaluated using the micromechanical solver. As shown in \cref{fig:pr1_inverse_noncovex_d}, the corresponding solver outputs closely match the reference data, including the local non-convex behavior.

\subsection{Adaptive Sampling}\label{sec:results:adaptive}

In this section, we demonstrate that the amount of data required to train a surrogate of the micromechanical solver can be significantly reduced by an adaptive sequential sampling strategy. Specifically, we compare the data efficiency of datasets generated by LHS and by the proposed sequential-search strategy in terms of dataset size and the resulting model accuracy.
Starting from $10$ initial samples drawn by LHS, the adaptive procedure sequentially appends new samples at each iteration for labeling, after which the surrogate is retrained by the ensembled relative MSE \cref{eq:loss_ensemble} until the stopping criterion is satisfied.
A detailed description of the settings in this work is provided in Appendix~\ref{apdx:hyperparams}.

We consider dataset sizes of $10$, $20$, $\dots$, $320$, and $640$, and evaluate each using a labeled reference dataset of size $10{,}000$ generated via LHS.
To assess prediction quality for each input parameter $\boldsymbol{\xi}$, we use the relative MSE \cref{eq:rel_MSE}.
All models are trained on their respective datasets using the same training protocol. The number of ensemble surrogates is set to $N_{\mathrm{ens}} = 3$.

After each retraining step, $1{,}000$ candidate samples are generated by LHS, and the total acquisition score \cref{eq:acquisition_tot} is evaluated for each candidate. The top $10$ samples with the largest scores are then labeled and appended to the dataset where we denote as $\mathrm{TopK}$ with $\mathrm{K}=10$. The adaptive sampling procedure is terminated once the dataset size reaches $640$, corresponding to a budget-based stopping criterion.
A schematic overview of the procedure is provided in Algorithm~\ref{alg:active_learning}.

\begin{figure}[htbp]
    \centering
    \begin{subfigure}{0.3\textwidth}
        \centering
        \includegraphics[width=\textwidth]{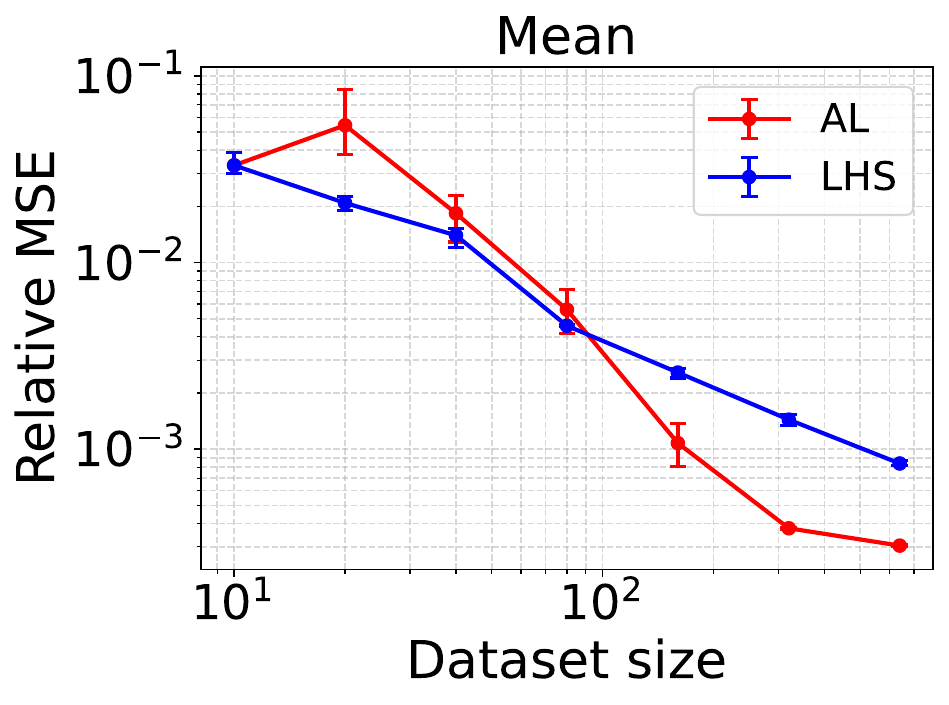}
        \caption{}\label{fig:pr1_active_learning_bench_a}
    \end{subfigure}
    \begin{subfigure}{0.3\textwidth}
        \centering
        \includegraphics[width=\textwidth]{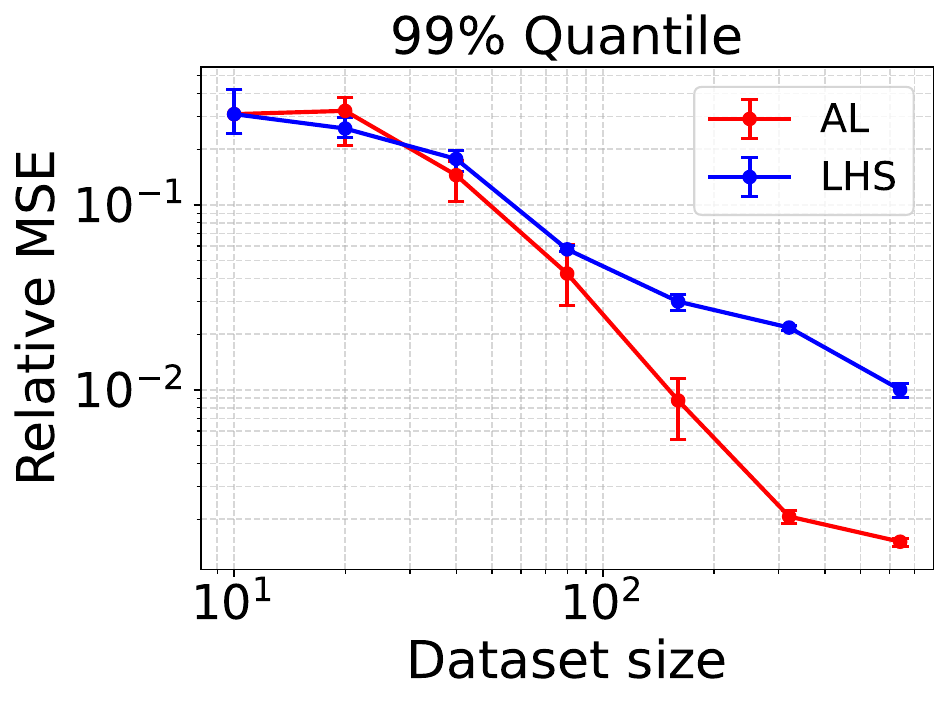}
        \caption{}\label{fig:pr1_active_learning_bench_b}
    \end{subfigure}
    \begin{subfigure}{0.3\textwidth}
        \centering
        \includegraphics[width=\textwidth]{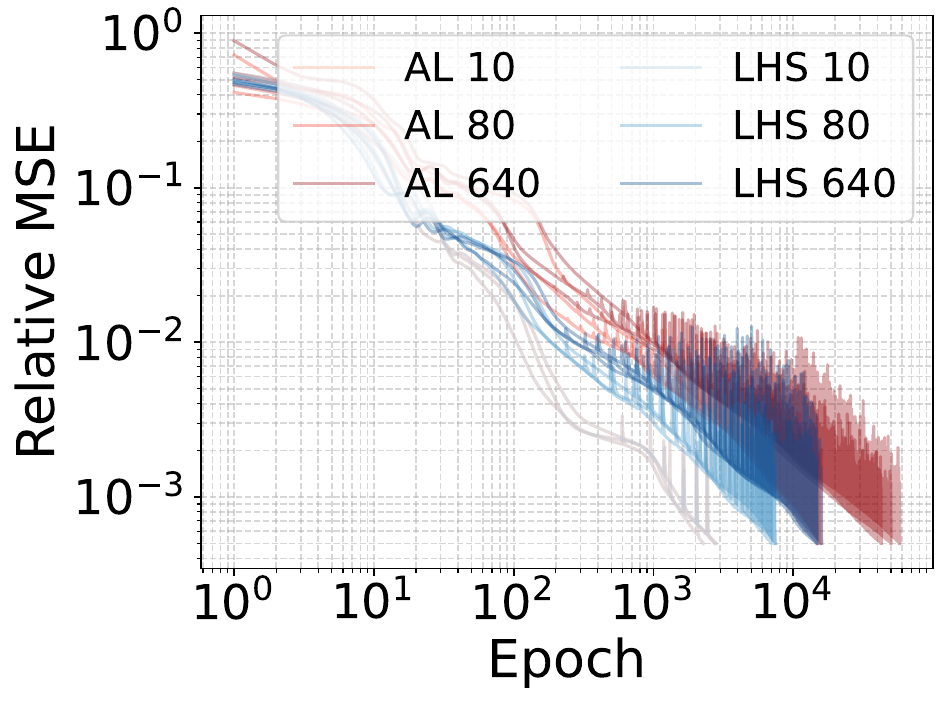}
        \caption{}\label{fig:pr1_active_learning_bench_c}
    \end{subfigure}
    \caption{
    Comparison of data efficiency between LHS and adaptive sequential sampling.
    Here, AL denotes datasets obtained using the adaptive sequential sampling algorithm.
    (\subref{fig:pr1_active_learning_bench_a}) shows the relative MSE of surrogate models trained on adaptively acquired data and on LHS data, reported as the mean across realizations.
    (\subref{fig:pr1_active_learning_bench_b}) reports the corresponding relative MSE in terms of the 99\% quantile across realizations.
    These statistics are averaged over three trials, with error bars indicating the minimum and maximum values across trials.
    (\subref{fig:pr1_active_learning_bench_c}) shows the loss histories of models trained on acquired datasets of sizes $10$, $80$, and $640$, with three trials shown for each dataset size.
    }
    \label{fig:pr1_active_learning_bench}
\end{figure}

The results of the adaptive sequential sampling are shown in \cref{fig:pr1_active_learning_bench}. \Cref{fig:pr1_active_learning_bench_a,fig:pr1_active_learning_bench_b} present a benchmark comparison of the data efficiency of LHS and the proposed adaptive sampling strategy for the neural operator. Overall, the errors evaluated on the reference dataset decrease as the dataset size increases. However, this trend is not monotonic in the small-size regimes; for example, the mean error increases from dataset sizes $10$ to $20$. Up to dataset sizes of $40$ and $80$, the gap between the two methods is not significant in terms of either the mean error or the $99\%$ quantile error. In fact, at sizes $40$ and $80$, the mean errors obtained with the active learning data are larger than those obtained with LHS. This behavior can be interpreted as a cold start problem in active learning: in the early-stage regimes, the model is often too biased for its uncertainty estimates to provide a reliable basis for query selection \cite{deng2018adversarial,maltz1995pointing}. A similar cold start issue has also been reported in computer vision \cite{chen2024making}, where many adaptive sampling schemes are outperformed by random baseline methods in such regimes.

However, beyond $160$, the difference becomes significant in both the mean and the $99\%$ quantile errors. Notably, the mean errors, averaged over three trials, obtained using $160$ adaptively selected samples are lower than those obtained using $320$ samples generated by LHS, and in one of the three trials, is even comparable to those obtained using $640$ samples generated by LHS. This trend becomes even clearer in the $99\%$ quantile errors, where the average error obtained using $160$ adaptively selected samples is slightly lower than that obtained using $640$ samples generated by LHS. \Cref{fig:pr1_active_learning_bench_c} shows the corresponding loss histories. For a fixed dataset size, the number of epochs required to reach a comparable relative MSE tends to be larger for adaptively sampled data than for datasets generated by LHS. This indicates that the adaptive sequential sampling algorithm preferentially selects samples that are more challenging for the model to learn. Overall, these results support that the proposed active learning strategy improves data efficiency by focusing on informative regions of the parameter space, leading to better predictive performance for a given data budget.

\section{Conclusion}\label{sec:conclusion}

In this work, we propose a physics-informed neural operator surrogate based on DeepONet that provides an efficient and differentiable framework for identifying microstructural parameters in granular micromechanics-based models, addressing the challenges of determining microstructure configurations that guarantee a target failure envelope. The formulation is designed to learn from and perform inference on failure envelopes sampled non-uniformly and represented as point clouds, enabling integration of data from different schemes, models, and experiments without discretization constraints. To enforce mechanical admissibility, a curvature-based regularization motivated by Drucker’s postulate is introduced. The results show that this regularization improves the physical plausibility of the predicted envelopes in both forward and inverse identification, with the trained model effectively acting as the closest convex projection of the queried failure envelope.

A computationally efficient approach based on finite difference approximations and batched evaluations is introduced to significantly accelerate training compared to common use of automatic differentiation. In addition, an adaptive sampling strategy based on ensembles of the proposed surrogate is developed, using an acquisition criterion that accounts for both predictive ambiguity and data diversity. The results demonstrate that this strategy improves data efficiency by preferentially selecting informative samples, particularly in regions associated with complex or extreme responses.

Overall, the proposed surrogate model provides an effective computational tool for both scientific analysis and data-efficient exploration of complex micromechanical response landscapes.

\section*{Data Availability}
All data used in this work are either publicly available or can be reproduced from the information provided and are available from the corresponding author upon reasonable request.

\section*{Acknowledgment}
B.B. acknowledges support from the startup fund provided by Northwestern University.

\section*{Declaration on the Use of Generative AI}

The authors used ChatGPT and Claude, a large language model developed by OpenAI and Anthropic, to assist with English language editing and grammar refinement. The scientific content, technical interpretations, and conclusions are solely the responsibility of the authors.

\bibliographystyle{plainnat}
\bibliography{bibliography}

\appendix

\section{Hyperparameters}\label{apdx:hyperparams}

Branch and trunk networks use MLPs with two hidden layers of width $32$ and hyperbolic tangent activation functions.  
Trunk network outputs 20 basis functions.
The DeepONet output is then mapped through a softplus function,
\begin{equation}
    \mathrm{Softplus}(x)=
    \frac{1}{\beta}
    \log\!\left(1+\exp(\beta x)\right),
\end{equation}
where the softplus sharpness parameter $\beta=5$ is used throughout.
All models are trained using the ADAM optimizer \cite{kingma2014adam} with a learning rate of $10^{-3}$.
The relative MSE loss \cref{eq:loss_rel_mse} is used for training, and the training is terminated once this loss falls below $5\times10^{-4}$.

For the systematic comparison of the differentiation schemes shown in \cref{fig:bench_training}, the model is trained for $30{,}000$ epochs with a curvature-penalty parameter of $\lambda=10$.

For the two surrogate models considered in Section~\ref{sec:results:surrogate}, a batch size of $512$ is used, so that each batch consists of $512$ parameter vectors $\boldsymbol{\xi}$. The curvature penalty parameter is set to $\lambda = 10$, which we found sufficient to promote convexity of the surrogate output.

For inverse identification, the optimization is terminated once the projected gradient residual $\mathcal{R}$ in \cref{eq:gradient_residual} falls below $10^{-8}$. The optimization is performed independently for each initial guess, without batching. After each step, $\boldsymbol{\xi}$ is clamped to $[-1,1]^p$ to remain consistent with the normalized input domain. The ADAM optimizer is used with a learning rate of $10^{-1}$.

Training is performed in a full-batch setting. To assess dataset informativeness in \cref{fig:pr1_active_learning_bench}, all models are initialized with the same set of random seeds across all datasets.

\section{Stress-Controlled Numerical Implementation}
\label{apdx:ForwardEuler}

The GMA developed in Section \ref{sec:GMA} is used to generate biaxial failure envelopes via simulating proportional loading paths in the biaxial stress space. 
In each loading path, stress tensor at the updated pseudo-time $\tau+\Delta \tau$ is defined as $\bm{\sigma}^{\tau+\Delta \tau} = \bm{\sigma}^{\tau} + \text{d}\bm{\sigma}$, where the biaxial stress increments are given as $\text{d}\sigma_{11}$ and $\text{d}\sigma_{22}$, with other stress increments equal to zero. 
Since the inter-granular constitutive laws defined in Section \ref{sec:data} are non-linear and path-dependent, the macroscopic tangent stiffness tensor $\mathbb{C}$ is a function of the current state of damage and displacement. However, since the loading is done in stress-controlled manner, the macroscopic tangent operator cannot be identified straightforwardly.

In this study, to solve the incremental constitutive equations without the iterative overhead of implicit methods, we utilize a forward Euler scheme.
Within the utilized forward Euler scheme, it is assumed that the tangent operator used for updating the strain tensor can be calculated at the pseudo-time step $\tau$.
To this end, the strain tensor in pesudo-time $\tau$ is used to find the local grain-scale kinematics via \cref{eq:dndw}. For each grain-pair orientation, the history variables are checked to determine if each interaction is in a state of active loading or unloading/reloading (separately for normal and tangential interactions). The local forces and tangent stiffness coefficients are then updated according to \cref{eq:fnfw-damage}. 
Finally, the macroscopic tangent is derived as $\mathbb{C}^{\tau} = \mathbb{C}\left(\boldsymbol{\epsilon}^{\tau}\right)$ via \cref{eq:Cijkl-int}. 
Given the incremental stress, the strain components at step $\tau+\Delta \tau$ are calculated as
corresponding macroscopic strain increment $\text{d}\bm{\epsilon}$ is obtained by inverting the tangent stiffness matrix, resulting in:
\begin{equation}
\epsilon_{ij}^{\tau+\Delta \tau} = \epsilon_{ij}^{\tau} + \mathrm{d}\epsilon_{ij}^{\tau}; \quad \text{where} \quad
\mathrm{d}\epsilon_{ij}^{\tau} = [C_{ijkl}^{\tau}]^{-1} \mathrm{d}\sigma_{kl}^{\tau}.
\label{e-update}
\end{equation}

This explicit integration continues along each radial path until the determinant of the tangent stiffness tensor $\mathbb{C}$ vanishes, at which point the stress state is recorded as a point on the failure envelope. 
To ensure numerical stability and to maintain the accuracy of the linearization in \cref{e-update}, small stress increments are utilized \cite{butcher2016numerical}. This explicit scheme is particularly efficient for the autonomous generation of the large datasets required for the machine learning framework, as it avoids the iterative computational overhead of implicit methods while directly providing the tangent operator used for failure detection.

\end{document}